\documentclass[a4paper,12pt]{article}

\usepackage{amsmath,amssymb,amsfonts}
\usepackage[dvips]{graphicx}
\usepackage{epsfig}
\usepackage{color}
\usepackage{verbatim}
\usepackage{calc}
\usepackage{bbm}
\usepackage{setspace}
\usepackage{array}
\usepackage{epstopdf}
\usepackage[caption=false]{subfig}
\usepackage[left=2.4cm,top=3.3cm,right=2.4cm,bottom=3.3cm,bindingoffset=0cm]{geometry}

\newcommand{\beq}{\begin{eqnarray}}
\newcommand{\eeq}{\end{eqnarray}}
\newcommand{\bea}{\begin{eqnarray}}
\newcommand{\eea}{\end{eqnarray}}
\newcommand{\be}{\begin{equation}}
\newcommand{\ee}{\end{equation}}

\def\brc{\langle}
\def\ckt{\rangle}

\def\de{\partial}

\def\1{\mathbbm{1}}

\def\Tr{{\rm Tr}}
\def\CN{{\cal N}}
\def\Im{{\rm Im}}

\def\c{\rm c}

\def\nc{\sigma}

\numberwithin{equation}{section}

\input{my_opt.sty}

\begin{document}

\title{
\begin{flushright}\ \vskip -1.5cm {\small {IFUP-TH-2018}}\end{flushright}
\vskip 20pt
\bf{ \Large Confinement and XSB in QCD :  \\  Mysteries and beauty of soliton dynamics in nonAbelian gauge theories \footnote{$CP^3$-Origins distinguished lecture, University of Southern Denmark, Odense, 12 March 2018  }}
}
\vskip 10pt  
\author{    Kenichi Konishi$^{(1,2)}$    \\[15pt]
{\em \footnotesize
$^{(1)}$Department of Physics ``E. Fermi'', University of Pisa}\\[-5pt]
{\em \footnotesize
Largo Pontecorvo, 3, Ed. C, 56127 Pisa, Italy}\\[3pt]
{\em \footnotesize
$^{(2)}$INFN, Sezione di Pisa,    
Largo Pontecorvo, 3, Ed. C, 56127 Pisa, Italy}\\[3pt]
{ \footnotesize kenichi.konishi@unipi.it,} 
}
\date{ March  2018}

\maketitle

\begin{abstract}
   Deep insights into the possible infrared dynamics of strongly-coupled nonAbelian gauge theories such as QCD come from the analyses of ${\cal N}=1$ or ${\cal N}=2$ supersymmetric gauge theories. Central in the whole discussion will be the topological soliton monopoles and vortices 
  and their quantum dynamics. We review the arguments  that nonAbelian monopoles,  free from the classic "difficulties",  can be defined semi-classically via the topology and stability connection to the better understood nonAbelian vortices.    Recent results on $CP^{N-1}$ models on $2D$ worldsheet of finite width, establish the quantum mechanical nature of such nonAbelian monopoles.  An interesting class of RG flows and emergence of confining vacua "nearby"  strongly-coupled infrared-fixed point  (IRFP) conformal theories are discussed in the context of most singular vacua in   ${\cal N}=2$ supersymmetric QCD.
  Certain analogy with the real-world QCD is drawn.   In many systems,  
   color-flavor locking emerges as a crucial mechanism for the gauge system to avoid dynamical Abelianization.

~~~~

  \end{abstract}

\tableofcontents
\newpage

\newpage

\section{Introduction}

In spite of many years of study, the true nature of quark confinement and chiral symmetry breaking in the real world of strong interactions, believed to be described by SU(3) Quantum Chromodynamics (QCD), still evades us. The main difficulty lies in our lack of understanding of the effective low-energy degrees of freedom and of their interactions.  As it turned out, deep insights into the possible types of infrared dynamics of strongly coupled nonAbelian gauge theories such as QCD, come from the analyses of supersymmetric models,  where many of the beautiful ideas born in earlier studies of topological features of quantum field theories are tested, clarified and enriched. Central in the whole discussion will be the topological soliton monopoles and vortices, especially their quantum dynamics, dualities and anomalies. 

Solitons first appeared in physics as solitary waves on the water;  there are many examples in Nature, from tornados in air and vortices of hydrodynamics, vortices in plasmas, 
and in more subtle setting of superconductor, as Abrikosov vortices \cite{Abrikosov} of magnetic fields.  Also various kinds of important soliton excitations 
appear in condensed matter physics,  in Quantum Hall effects, BE condensed cold atoms, Neutron stars, and so on.  

In the context of physics of the fundamental interactions the first to appear was the Dirac monopole \cite{Dirac};  many years later it was realized that the soliton monopoles (and vortices) appeared as topologically stable finite-energy field configurations in spontaneously broken gauge theories \cite{TP, NO}.  These works opened whole areas of applications in quantum field theories, such as instantons, skyrmions, domain walls, lumps, etc. \cite{RebbiSoliani}.   Especially interesting among these in the context of quark confinement are the nonAbelian monopoles which, though discovered  \cite{GNO}   soon after 't Hooft's and Polyakov's papers (the nonAbelian {\it vortices} have been found much later \cite{Hanany:2003hp,Auzzi:2003fs,Shifman:2004dr}), are still somewhat covered by mysteries. They are likely to hold the key to proper understanding of the confinement and chiral symmetry breaking in QCD.

 \section{Magnetic monopoles}
 
 \subsection{From classical to quantum magnetic monopoles}
 
 The whole story started from the observation   that the vacuum Maxwell equations are invariant under the electromagnetic duality transformations, ${\bf E} \to {\bf B}, \quad {\bf B} \to -{\bf E}$. This invariance is broken in Nature by the presence of electrically charged sources but not the magnetic ones. Dirac \cite{Dirac}  however has shown that in principle there is nothing that forbids the existence of magnetically charged particles in Nature, as long as the famous quantization condition is obeyed which involves the product of electric and magnetic charges,
 \be    g g_m = \frac{n}{2}\;, \qquad  n=0, \pm 1, \pm 2, \ldots.  
 \ee
 It is interesting to note that from the beginning it was clear that such a quantization condition was related to the topology - or homotopy - of the mapping from the space to the 
 internal, gauge group space \cite{WY}.  Actually his quantization condition can be generalized to
  \be  g^{(1)} g_m^{(2)} -   g^{(2)} g_m^{(1)}   = \frac{n}{2}\;,
 \ee
allowing for the presence of dyons, carrying both electric and magnetic charges,   $(g^{(1)}, g_m^{(1)})$ and $(g^{(2)}, g_m^{(2)} )$.  Dirac's quantization condition could explain why the electric charges in Nature are quantized in the unit $g=e$,  the electric charges of the electron and the proton. 
 
 Many years later,  't Hooft and Polyakov have shown \cite{TP} that the magnetic monopoles appeared in spontaneously broken gauge theories as topologically stable, finite-energy  soliton-like classical configurations (solutions of the equations of motion).  They look like,  in an $SU(2)$ theory broken to $U(1)$ by an adjoint scalar VEV $\phi$, 
 \be    A_i({\bf r}) =   \epsilon_{aji} \frac{r^j}{r^2}  a(r)\, S^a  \;,  \qquad  \phi({\bf r}) = v_1 \, \frac{ r^a  S^a}{r}  \chi(r)\;,\qquad  \chi(r), a(r) \stackrel {r \to \infty}  {\longrightarrow} 1
 \ee
 where $S^a$, $a=1,2,3$ are the $SU(2)$ generators.  The functions $a(r)$ and $\chi(r)$ satisfy simple coupled equations following from the YM  field equations.

 Apart from direct interest in observing experimentally such objects, as expected in grand unified theories, their work immediately raised the issue of the quantum properties of the soliton monopoles. 
 The problem was first formulated as the study of quantization of the gauge and matter fields  in the classical soliton monopole background. This investigation has led to many beautiful and subtle results such as charge fractionalization \cite{GW}, Jackiw-Rebbi effect \cite{JR}, Witten effect \cite{Witten1}, Rubakov effect \cite{Rubakov}, wrong statistics of monopole fermion composites \cite{HHGH}, and so on, around the years  '74-'80 \cite{RebbiSoliani}.  Some puzzles remained such as how to put together two monopoles,  how the theory gets renormalized in the presence of an electron and a monopole,  with coupling constants running towards  opposite directions, and so on.
 
 A breakthrough in the study of quantum dynamics of monopoles came with the discovery of the exact Seiberg-Witten solutions, in ${\cal N}=2$ supersymmetric gauge theories \cite{SW1,SUN}.  The ${\cal N}=2$ $SU(2)$ gauge theory  is described by the fields 
 \be   W =  (A_{\mu}, \lambda), \qquad  \Phi = (\phi, \psi)\;,
 \ee
 where $\lambda$ is the gauge fermion,  $\phi$ and $\psi$ are the complex scalar and another Weyl fermion, all in the adjoint representation. The presence of the two Weyl fermions with the same properties shows that the theory is invariant under a global $SU_R(2)$ symmetry, which is the basis for the ${\cal N}=2$ supersymmetry itself.  A characteristics of the theory is the presence of the classical moduli space (the flat direction) of vacua, 
 \be  \brc  \phi  \ckt =  \left(\begin{array}{cc}a & 0 \\0 & -a\end{array}\right)\;,      \qquad    u \equiv  \Tr \, \Phi^2\;, \qquad  \quad a, u  \in  {\mathbf C}\;. 
 \ee
 As a result of ${\cal N}=2$ supersymmetry, the low-energy effective action (to the lowest number of the field derivatives)  has the structure, 
 \be   {\cal L}_{eff} =  \Im  \, \int d^4 \theta \, {\bar A} \, \frac{\de F(A)}{\de A} +  \int d^2\theta \,\frac{\de^2 F}{\de A^2}  W^{\alpha} W_{\alpha} 
 \ee
 where a holomorphic function $F(A)$ is known as prepotential.   $A_D =   \tfrac{\de F(A)}{\de A} $  is the dual coordinate, 
 \be  \tau =  \frac{\de^2 F}{\de A^2} = \frac{d A_D}{d A} =  \frac{\theta_{eff}}{2\pi} + \frac{4\pi i}{g_{eff}^2}
 \ee
describes the low-energy effective $\theta$ parameter and the coupling constant.  It turns out that the structure of the effective action above is  (formally) invariant under the (generalized) Legendre transformations  - $SL(2, Z)$:
\be \left(\begin{array}{c}A_D \\A\end{array}\right)   \to    \left(\begin{array}{cc}a & b \\c & d\end{array}\right) \,  \left(\begin{array}{c}A_D \\A\end{array}\right)\;, \qquad  ad- bc =1\;. 
\ee
\be     \left(\begin{array}{c}   \frac{ \delta L}{\delta F^+_{\mu \nu} }   \\   F_{\mu \nu}  \end{array}\right)  \to  \left(\begin{array}{cc}a & b \\c & d\end{array}\right) \,   
 \left(\begin{array}{c}  \frac{ \delta L}{\delta F^+_{\mu \nu} }  \\   F_{\mu \nu}    \end{array}\right)           \;,   \qquad    F^+_{\mu \nu} = F_{\mu \nu}+ i {\tilde F}_{\mu \nu} 
\ee
Which basis is to be used depends on the vacuum, $u$.   When $|u| \gg \Lambda^2$, the physics is semiclassical, and  the original electric variables $A,  F_{\mu \nu}$
are the appropriate variables.  Near  $u=  \Lambda^2$,  by assumption, a magnetic monopole becomes light,  so the appropriate variables are  $A_D,  F_{D\,  \mu \nu}$,
obtained by the duality transformation {\footnotesize $\left(\begin{array}{cc}a & b \\c & d\end{array}\right)=\left(\begin{array}{cc}  0 & 1 \\  -1 & 0  \end{array}\right).$}
The known magnetic charge appearing at the vacuum  $u=\Lambda^2$, $(n_m, n_e)=(1,0)$,  then  determines \cite{SW1} the prepotential $F(A)$.  

  In particular, the Seiberg-Witten curve  (which defines a complex one-dimensional surface as  $x, y \in {\mathbf C}$) ,
  \be   y^2=   (x-u) (x-\Lambda^2) (x+ \Lambda^2)     \label{curve}
  \ee
  solves the theory, in the sense that it determines 
  \be    \frac{d A_D}{du} = \oint_{\alpha} \frac{dx}{y}\;, \quad   \frac{d A}{du} = \oint_{\beta} \frac{dx}{y}\;  
  \ee
  and hence the prepotential $F(A) $  itself,  
  where $\alpha$ and $\beta$ are the two canonical cycles on the  auxiliary torus on which the variable $x$ is defined. 
  It is truly remarkable that  the simple curve (\ref{curve}) encodes  all perturbative and nonperturbative (instantons) quantum corrections exactly. 
  

By assumption, near $u=\Lambda^2$, a  $(n_m, n_e)=(1,0)$ monopole becomes light \footnote{At the other singular vacuum,  $u=-\Lambda^2$  a $(n_m, n_e)=(1,1)$ dyon appears as massless matter field. },  so the effective action must be complemented as
\be    L_{eff} (A_D, F_D^{\mu \nu}) + \int   d^4\theta \, {\bar M}  e^V M  +  (M \to {\tilde M}) + \sqrt{2} \int d^2 \theta  \, M  A_D {\tilde M} \;,\label{effSW}   \ee
  the new terms describe the light magnetic monopoles coupled minimally to the dual gauge fields. 

  The Seiberg-Witten curves can be written for other gauge groups as well. For instance, for $SU(N)$ theory with $N_f$ quarks,  it reads \cite{SUN}
  \be   y^2=   \prod_{i=1}^{N} (x- \phi_i)^2 - \Lambda^{2N-N_f} \prod_{a=1}{N_f} (x+ m_a)\;;
  \ee
 and  for $SO(N)$ theory with $N_f$ quarks in the vector representation, 
   \be   y^2= x   \prod_{i=1}^{[N/2]} (x- \phi_i^2)^2 -  4\Lambda^{2(N-N_f)}  x^{2 +\epsilon} \prod_{a=1}^{N_f} (x+ m_a^2)\;;\qquad \epsilon=\pm1 \;.
  \ee
Reviews on the magnetic monopoles can be found in \cite{Coleman50,Konishi75}.

\subsection{Monopole condensation and confinement}

Coming back to $SU(2)$ theory,  it was shown \cite{SW1} that a small ${\cal N}=1$ perturbation, 
\be    \Delta L =    \mu \Phi^2 |_{\theta^2}  =  -\mu \, \psi \psi + \ldots     \label{N=1pert}
\ee
which gives mass to the fields $(\phi, \psi)$ and breaks supersymmetry to ${\cal N}=1$,  induces, at low energies,  a modification of the effective action
\be    \Delta L_{eff}  =    \mu \,  u(A_D)  \simeq     \mu \,   (\Lambda^2 +  \frac{\de u}{\de A_D} A_D + \ldots )\;,
\ee
to (\ref{effSW})  such that  the equation of motion of the low-energy theory now gives
\be   A_D=0\;; \quad  ( u = \Lambda^2)\;;  \qquad     \brc M \ckt =  \sqrt{ \mu \,  \frac{\de u}{\de A_D} |_{A_D=0} }\sim  \sqrt{ \mu \,  \Lambda}\,.
\ee
The magnetic monopole has condensed (dual Higgs), and at the same time the continuous vacuum degeneracy has been eliminated
by the pertubation, leaving only two vacua, $u=\pm \Lambda^2$.   The theory possesses a vortex string solution, and according to Nambu, Mandelstam and 't Hooft, this is a confining vacuum.   This may be termed dual superconductor mechanism of confinement. 
Importance of this work \cite{SW1} is that  for the first time ever the occurrence of confinement has been proven analytically in $4D$ Yang-Mills theory. 

Actually, this brings us to a little puzzle. The low-energy theory is a (dual) Abelian Higgs model.  The monopole condensation means that the system 
generates vortices,  with a  quantized integer winding number, 
\be  \pi_1(U(1)) = {\mathbf Z} \;.
\ee
The underlying $SU(2)$ theory however does not allow for a vortex of arbitrary winding number.  As the matter fieds $\lambda, \phi, \psi$ are all in the adjoint representation, the gauge group is actually 
\be  \frac{SU(2)}{{\mathbf Z}_2} \sim SO(3)\;, 
\ee
and the allowed charges (at which the vortices can end) or the winding number,  are classified by  \cite{WY}
\be   \pi_1( \frac{SU(2)}{{\mathbf Z}_2} ) = {\mathbf Z}_2\;. \label{Z2}
\ee

This implies first that the meaning of confinement in the Seiberg-Witten theory is that a fundamental chromoelectric charge  (a particle in the doublet representation of the $SU(2)$ gauge group - or simply a "quark"), {\it if introduced in the theory}, would be confined by the vortex string in the  $\brc M \ckt \ne 0$ medium. This is similar to the standard idea of confinement in pure $SU(N)$ gauge theory.  How exactly confinement of such a fundamental charge occurs is not known.  In the context of ${\cal N}=2$ theory, on the other hand, it is well-known what happens when fields in the fundamental representation are introduced \cite{SW1}. There must be a way to use such a knowledge to illustrate better the meaning of quark confinement. 

Secondly,  a doubly wound vortex of the low-energy Seiberg-Witten $SU(2)$ theory, when a second order effect in $\mu$ is taken into account, turns out to be stable against decay to two fundamental vortices, and 
the system would produce exotic hadrons, quite unlikely the standard QCD \cite{YungVain}.  As the gauge group of the system has the property $ \pi_1(\frac{SU(2)}{{\mathbf Z}_2}) = {\mathbf Z}_2 $, a doubly wound vortex  
could eventually decay by the pair production of massive $W$ bosons from the vacuum, but these nonperturbative processes are highly suppressed,  if
$\Lambda \gg \mu$. 
 Again, it should be possible to clarify better how these processes occur.

 \subsection{'t Hooft's  $({\mathbf Z}_N,  {\mathbf Z}_N)$ classification }
 
 In passing let us recall that in pure $SU(N)$ Yang-Mills theory  (supersymmetry is irrelevant here), where the gauge group is ${SU(N)}/{{\mathbf Z}_N}$,   the  allowed "electric" and "magnetic"   charges 
 are both  classified by the center ${\mathbf Z}_N \subset SU(N)$. This is simply due to the fact that any  interactions involving gluons cannot change the 
 $N$-ality charge,  on the one hand, and the fact that 
 \be   \pi_1(\frac{SU(N)}{{\mathbf Z}_N}) = {\mathbf Z}_N
 \ee 
 tells us \cite{WY} that the possible magnetic charges are quantized by ${\mathbf Z}_N$.  Thus the possible vacua   in pure  $SU(N)$ YM can be  classified by the 
 \be  ({\mathbf Z}_N,  {\mathbf Z}_N)  
 \ee
 charge of the entity which condenses.  If $(a, b)$ field is condensed, a particle $X=(A,B)$ where 
 \be  \brc x, X \ckt=  a B- b A \ne  0, \qquad mod \,\, N 
 \ee
 is  confined.

  \subsection{Monopole condensation and confinement in a non supersymmetric vacuum}

\begin{figure}[h]
\begin{center}
\includegraphics[width=4in]{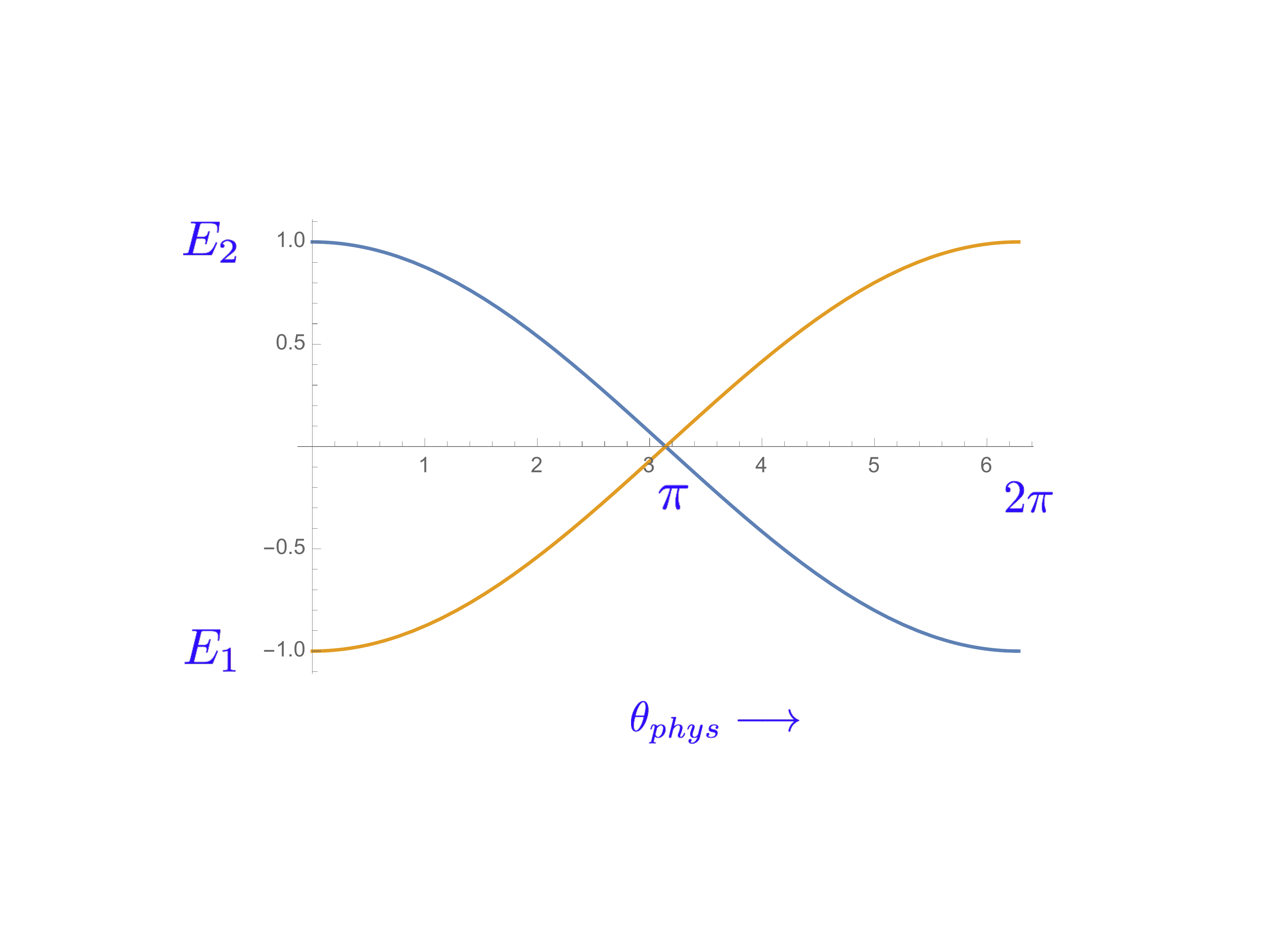}
\caption{ }
\label{2vacua}
\end{center}
\end{figure}

  There is no difficulty in constructing models without supersymmetry, in which monopole condensation and confinement can be studied analytically \cite{EHS,KK}.  The idea is to set up 
  ${\cal N}=0$ perturbation  (the gaugino mass)  
  \be   m_{\lambda} \lambda \lambda  + h.c.
  \ee
  together with somewhat larger  ${\cal N}=1$ perturbation, Eq.~(\ref{N=1pert}).  The remaining degeneracy between the two vacua is now lifted, and the energy of the two vacua oscillates
  as  a function of the physical $\theta$ parameter, 
  \be   \theta_{phs} =     \theta + 2 ({\rm Arg} \mu + {\rm Arg} m_{\lambda})\;,  
  \ee 
  \be  \Delta E =   -  m_{\lambda} \brc \lambda \lambda \ckt +c.c. = - \frac{16 \pi^2 \mu \,m_{\lambda}}{g^2}  \brc \Phi^2 \ckt +c.c =  \mp \, \epsilon \,\cos ({\theta_{phys}}/{2})\;,
  \ee
  where $\epsilon=  \frac{16 \pi^2 | \mu \,m_{\lambda}   \Lambda^2| }{g^2} $.    See Fig.~\ref{2vacua}.
 
  \section{Magnetic monopole condensation and confinement in QCD}
  
  What about the real world QCD?    't Hooft, Nambu and Mandelstam proposed around '80 that the QCD vacuum might be an Abelian dual superconductor, in which the gauge group is dynamically 
  reduced as 
  \be  SU(3) \to     U(1)^2  \to {\mathbf 1}\;. 
  \ee
  Many investigation followed, especially in the context of lattice approach,  but no convincing evidence emerged. 
  There are actually a few problems.  The fact that 
  \be  \pi_1 (U(1)^2)  =   {\mathbf Z}\times {\mathbf Z} \;, 
  \ee
  implies that the meson spectrum must be doubled, which is not observed in Nature. 
  Also, if confinement and chiral symmetry breaking are both induced by the same monopole condensation,  it would imply that a monopole carries 
  $({\underline  {N_f}},{\underline  {N_f}}^*) $ of   $ SU(N_f)_L \times SU(N_f)_R$ chiral symmetry charges,  $M^i_j$,  whose diagonal condensate
  \be  \brc M^i_j \ckt \sim  \Lambda \delta^i_j  \ee
    would induce the correct symmetry breaking,   
  \be   SU(N_f)_L \times SU(N_f)_R  \to  SU(N_f)_V\;.
  \ee
  But such a low-energy theory would actually possess an accidental $SU(N_f^2)$ global symmetry,  which implies a larger number of Nambu-Goldstone bosons, 
  $ \sim  N_f^4 - N_f^2 $, which, again,  is not observed in Nature. 
  
  If the gauge group does not Abelianize, i.e., if the gauge symmetry breaking occurs as
  \be    SU(3) \to \frac{SU(2) \times U(1)}{{\mathbf Z}_2}\;, \label{nonAbreaking}
  \ee
  then the problem of the meson spectrum doubling is naturally solved,  as  
  \be  \pi_1 (\frac{SU(2) \times U(1)}{{\mathbf Z}_2})  =   {\mathbf Z}\;.  
  \ee
  The problem of too-many-NG-bosons might also find a solution in such a context.\footnote{In the $r$-vacua of softly broken ${\cal N}=2$, $SU(N)$ theory with $N_f \ge 3$, 
  this potential difficulty is avoided \cite{APS,CKM}  by producing non-Abelian monopoles carrying the fundamental $SU(N_f)$ charge, rather than 
  Abelian monopoles in  ${N_f \choose  r}$-dimensional representation of $SU(N_f)$.}
 The problem with (\ref{nonAbreaking}) is that the nonAbelian monopoles arising from such a dynamical gauge symmetry breaking, are expected to be strongly coupled at low energies,  leaving aside possible problem with the very concept of nonAbelian monopoles in general (see below).  In any event, one would have a system in which both electric (quarks and gluons) and magnetic
 (magnetic monopoles and dual gauge bosons) variables become strongly coupled towards the infrared.

\section{Renormalization-group flow, infrared-fixed-points and confinement}

\begin{figure}
\begin{center}
\includegraphics[width=5in]{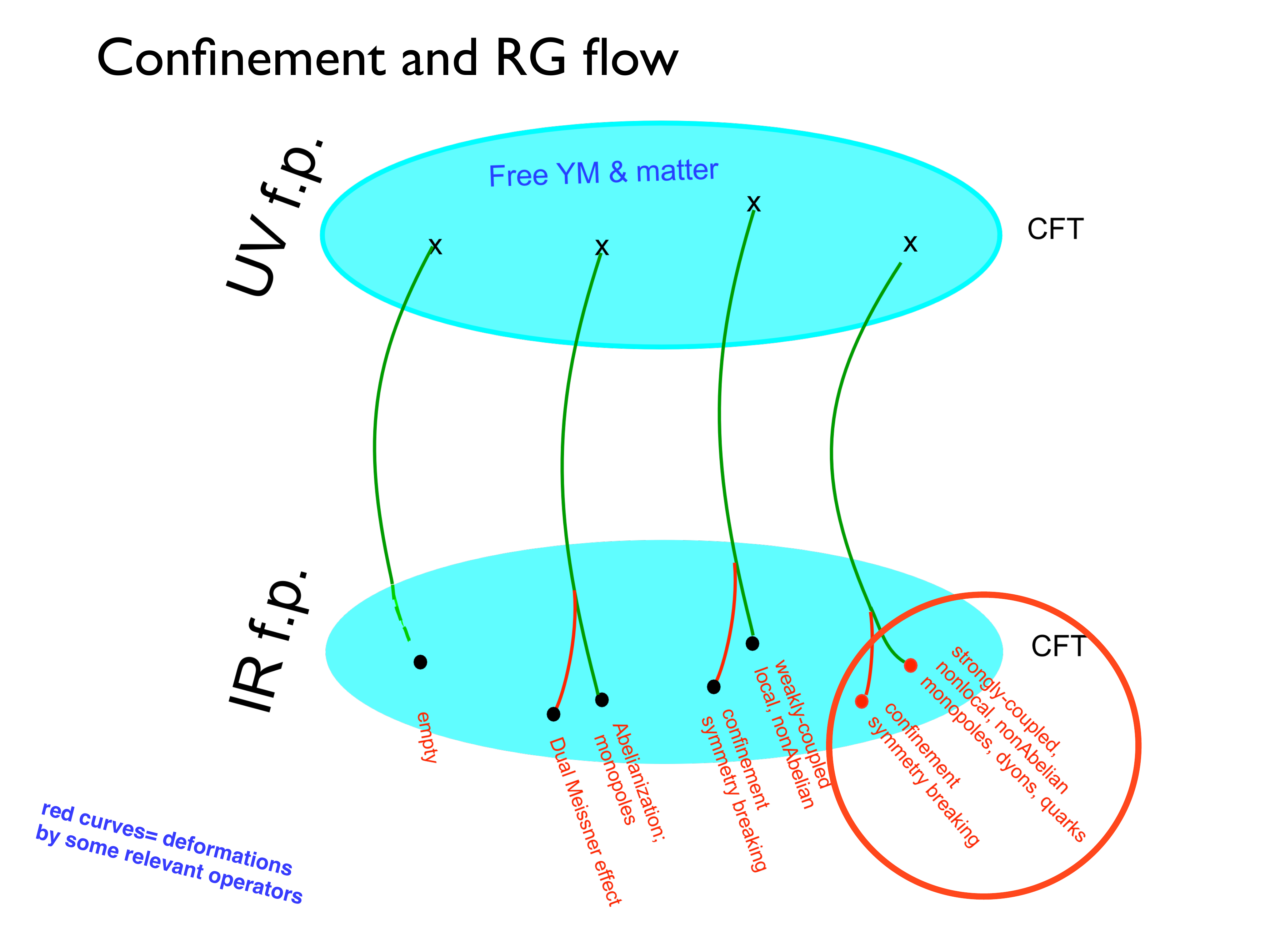}
\caption{ }
\label{RGflow}
\end{center}
\end{figure}

The above two candidate dynamical scenarios for QCD  are just some special cases of the possible remormalization-group flow in general asymptotically free theories. 
Consider a theory with some vacuum parameters,  or simply a space of theories,  which all become strongly coupled in the infrared. A theory may simply dynamically get Higgsed, the spectum is massive, and the infrared theory is empty. Or the infrared theory could be an Abelian or nonAbelian dual gauge theory of
weakly coupled monopoles, with their condensation leading to a confining vacuum. The most interesting possibility is that the infrared fixed point is a 
strongly-coupled theory of monopoles and dyons.  In Figure \ref{RGflow} the red curve denotes a deviation of the RG flow caused by a perturbation by some relevant operators, 
which may be introduced in the UV theory, or may be generated dynamically by the system itself. The figure illustrates the meaning of an apparently paradoxical idea that a confining vacuum is "close to an infrared-fixed-point conformal theory".

\begin{figure}
\begin{center}
\includegraphics[width=5in]{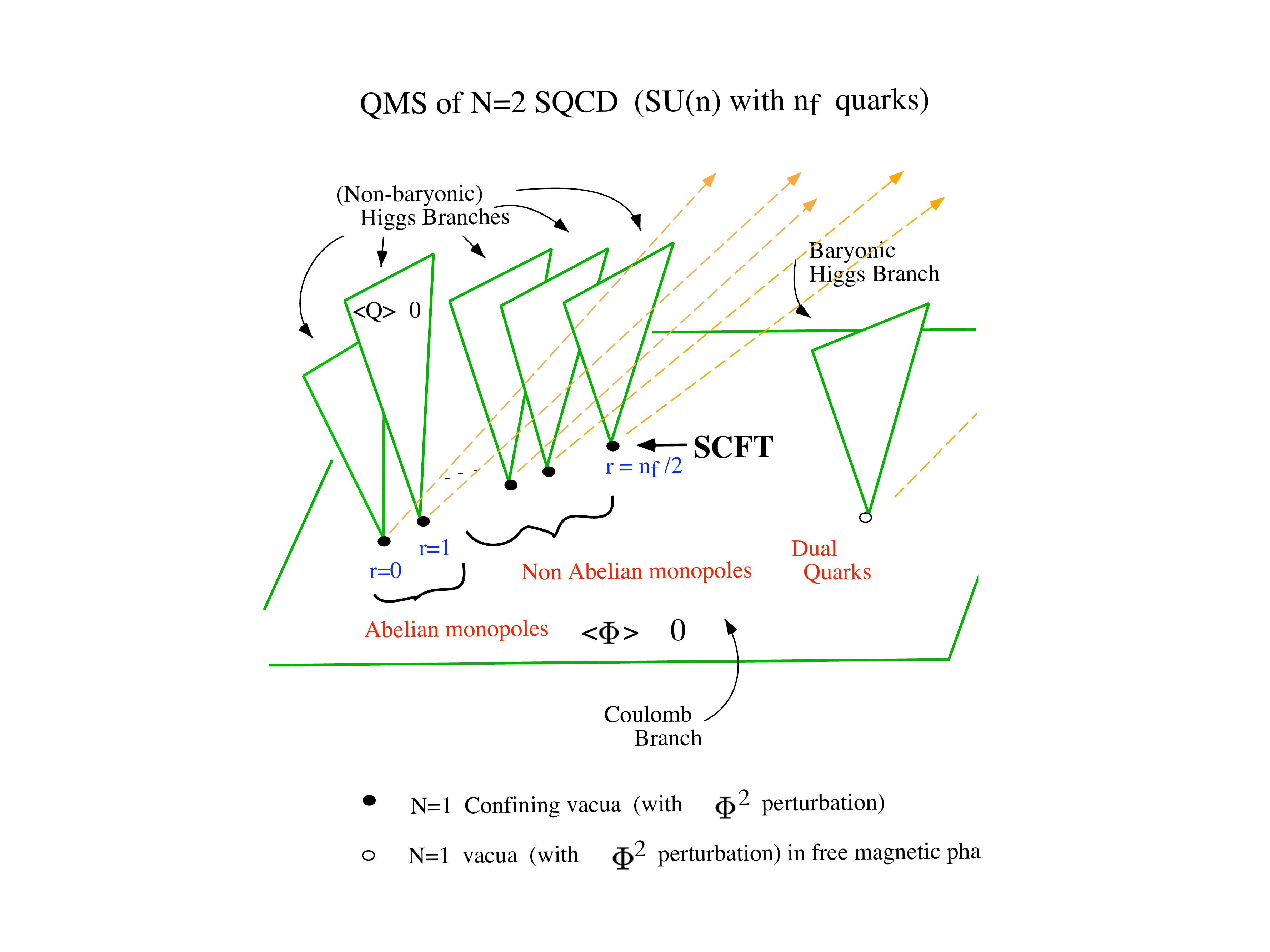}
\caption{ }
\label{SQCD1}
\end{center}
\end{figure}

\begin{figure}
\begin{center}
\includegraphics[width=5in]{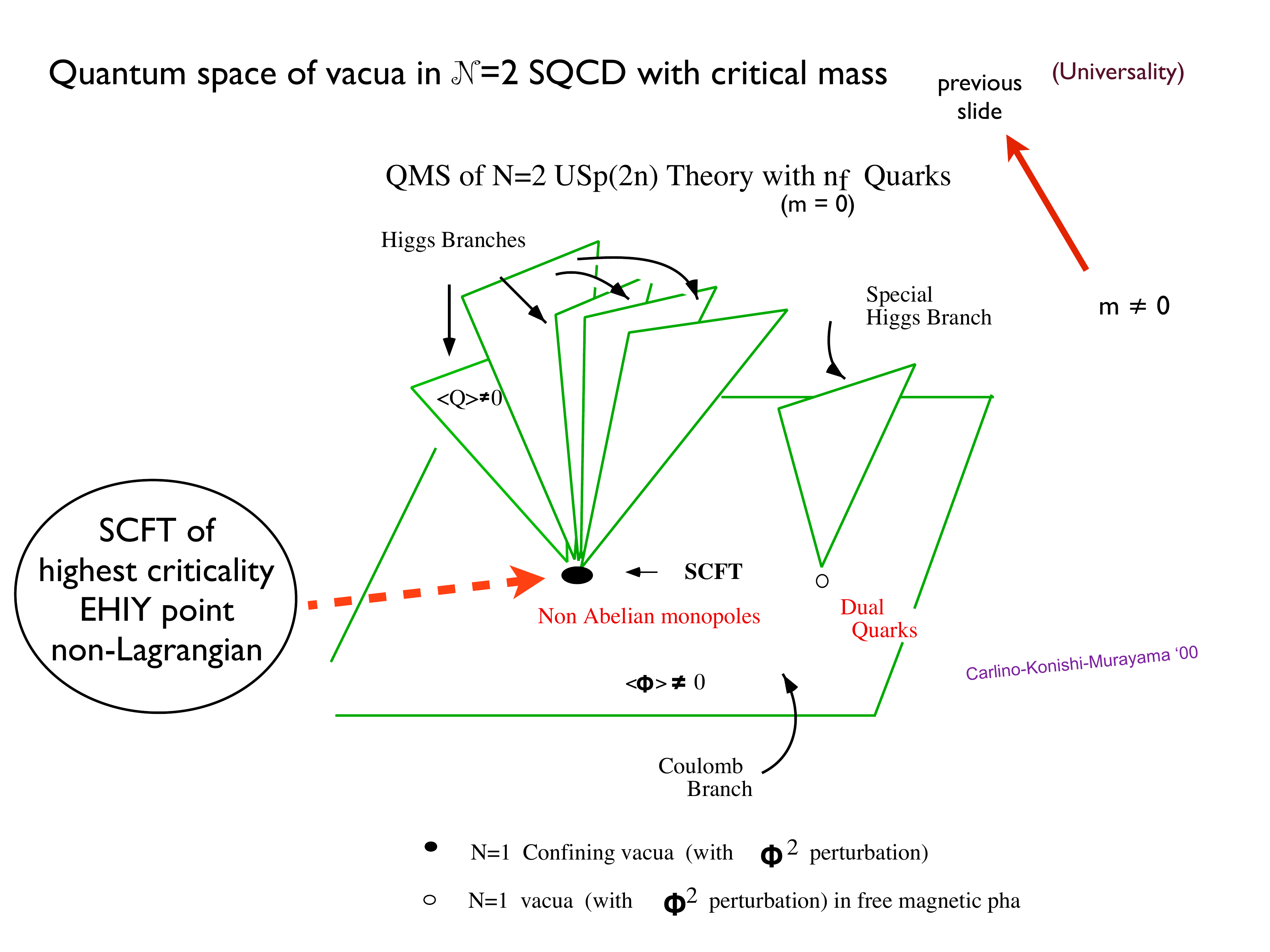}
\caption{ }
\label{SQCD2}
\end{center}
\end{figure}

A schematic view of possible vacua in softly broken ${\cal N}=2$  SQCD  with $SU(N)$ gauge group with $N_f$ flavors of quark multiplets 
\cite{CKM}
 is shown in Fig.~\ref{SQCD1}. 
It was only in 2011 that it was realized by Di Pietro and Giacomelli  \cite{SimoneLorenzo}  that all the $r$ vacua coalesce and form a singular EHIY vacuum \cite{AD}-\cite{Eguchi}, when the equal quark masses take a critical value, of the order of $\Lambda$.  See Fig.~\ref{SQCD2}.   On the other hand it was known from earlier analyses \cite{Eguchi} that the vacua of the $USp(2N)$ (or $SO(N)$) theory with $N_f$ quarks with $m=0$ are automatically such a strongly-coupled EHIY vacuum. Universality of the infrared-fixed-point SCFT's relate them.   We shall be interested below in the fate of these singular vacua,  when small ${\cal N}=1$ perturbation is added in the theory. 

Let us recall that the degrees of freedom of the (non collapsed) individual $r$ vacuum are  the nonAbelian monopoles (and some Abelian monopoles) carrying the 
dual $SU(r)$ gauge group charge, as well as the fundamental $SU(N_f)$ charge of the original UV theory \cite{APS,CKM}. 

\section{NonAbelian monopoles}

\subsection{$GNO$ quantization and classic difficulties}

Thus ${\cal N}=2$ supersymmetric gauge theories have no difficulties in generating nonAbelian monopoles as low-energy degrees of freedom.  
It is thus a highly nontrivial problem to understand  how the difficulties related to the classical notion of nonAbelian monopoles are overcome.   Semiclassically the nonAbelian monopoles appear when a gauge group $G$ is broken to a nonAbelian subgroup $H$, such that one finds
a collection of 't Hooft-Polyakov monopoles of a degenerate mass. Na{\"i}vely these are "related" by the unbroken group $H$.  The monopole field takes the form
\be    F_{ij} =  \epsilon_{ijk} \frac{r^k}{r^3} \, (\beta \cdot {\mathbf T})\;, 
\ee
where  $\beta$ is a constant vector  and ${\mathbf T}$  are the Cartan subalgebra generators of $H$.  One finds 
the "GNO" quantization rule \cite{GNO,EW},  
\be     2  \, \alpha \cdot \beta  \in {\mathbf Z}\;,  \label{GNO}
\ee
which is a natural generalization of the Dirac quantization condition, and which indeed arises in a very similar way.   By a known  group theory theorem, 
(\ref{GNO}) implies that the nonAbelian monopoles are labelled by weight vectors of the group ${\tilde H}$, the dual of $H$ (Table~\ref{GNOduals}) and not of $H$ itself \cite{GNO,EW,Harvey,ABEKM}.

Though such a classical construction of set of degenerate monopoles is straightforward, a more careful analysis reveals some problems, which have been coined in past by some even as a  "no-go" 
theorem. The first of these goes under the name of topological obstruction \cite{CDyons}:  as in the regular gauge the direction of the scalar field  in the monopole solution winds around in $SU(3)/U(2)$ at various space directions
$S^2$   (which corresponds precisely to nontrivial elements of $\pi_2(SU(3)/U(2))\sim \pi_1(U(2))\sim {\mathbf Z}$),  the embedding of  "unbroken" $SU(2)$ inside $SU(3)$  depends on ${\hat {\bf r}}$.  
One finds that it is not possible to define $SU(2)$ generators which are well defined in all directions \footnote{ If one works in the so-called 
singular gauge where the scalar field has a fixed orientation in $SU(3)$ asymptotically in all space directions  $\hat {\bf r}$, the problem manifests itself by the appearance of a Dirac string singularity in the gauge potential.}.  Another problem, which is an infinitesimal version of this same difficulty, is that when one tries to quantize the gauge fields in the background of the above mentioned "monopole" solution,  it is found that the gauge zeromodes are not normalizable in $3D$ \cite{DFHK,EW}.  One cannot therefore set up the quantized fields in the standard way as a sum of creation and annihilation operators associated with various (zero and nonzero) modes. Either way, trying to "rotate" the monopole solutions in the direction of the "unbroken $SU(2)$" would require an infinite amount of energy. 

The real issue, or the solution, is what GNO had already found out:  the monopole solutions are characterized by the weight vectors of the dual group ${\tilde H}$ and not of $H$.  ${\tilde H}$ is a group generated by nonzero root vectors,
\be   \alpha^* =   \frac{\alpha}{\alpha\cdot \alpha} \;,
\ee
where $ \alpha$ are the nonzero root vectors of $H$. See Table~\ref{GNOduals} for a few examples of dual pairs of groups, $H$ and ${\tilde H}$.
 The dual ${\tilde H}$ transformations are nonlocal field transformations in terms of the original gauge and matter fields, as this is a generalization of the electromagnetic duality. Thus, strictly speaking, those notorious classic difficulties are nonissues \footnote{In other words, the questions were not formulated correctly.   The problem is not how a monopole solution transforms under $H$ but under ${\tilde H}$. Except for some special cases such as $H=U(N)$,  the dual group ${\tilde H}$ is a different group: see Table~\ref{GNOduals}. More importantly the dual group ${\tilde H}$ does not act on the monopole solution as global field transformations as $H$.   See further discussions at the end of the next Section.}.

\begin{table}
  \centering 
  \begin{tabular}{ c |  c }
\hline
   $H$  &  ${\tilde H}$  \\
   \hline
   $U(N)$  &  $U(N) $ \\
$SU(N) $  &  $SU(N)/{\mathbf Z}_N$  \\
  $SO(2N)$  &  $Spin (2N)$   \\
   $SO(2N+1)$    &  $USp(2N)$  \\
\hline
\end{tabular}
  \caption{ }\label{GNOduals}
\end{table}

In any case, we {\it  know} that in the context of ${\cal N}=2$ theories nonAbelian monopoles are ubiquitous.  They appear typically in the singular points of quantum moduli space (where Abelian vacua collide),  and play the role of the low-energy degrees of freedom of the theory. There must be a way to understand how the theory produces them.  

\subsection{Defining nonAbelian monopoles via nonAbelian vortices}

A hint for the semiclassical origin of the nonAbelian monopoles  comes from studying hierarchical gauge symmetry breaking, in the presence of color-flavor locked symmetry,
\be   G \stackrel{\brc \phi\ckt = v_1}{\longrightarrow}  H   \stackrel{\brc q \ckt = v_2}{\longrightarrow}  {\mathbf 1}\;, \qquad  v_1 \gg v_2\;. 
\ee
For instance we may consider the breaking
 \be
SU(N+1)_{\rm color} \otimes SU(N)_{\rm flavor} 
  \stackrel{v_{1}}{\longrightarrow} 
(SU(N)\times U(1))_{\rm color}  \otimes SU(N)_{\rm flavor} 
\stackrel{v_{2}}{\longrightarrow} SU(N)_{C+F} \,.   \label{SUNhierarchy}
\ee
The idea is that 
as the exact symmetry  $SU(N)_{C+F} $ of the vacuum is broken by individual vortex as, 
\be   SU(N)_{C+F}   \to   SU(N-1)\times U(1)\;: 
\ee
the low-energy theory has vortex solutions which carries orientational zeromodes in
\be   CP^{N-1} \sim  \frac{SU(N)_{C+F}  }{SU(N-1)\times U(1)}\;. 
\ee
As the full theory cannot have vortices ($\pi_1 (SU(N+1)) = \emptyset$), such fluctuations must end  and be absorbed
by the monopoles at the extrema. But  then this mechanism endows the endpoint monopoles with the same $CP^{N-1}$  fluctuating degrees of freedom, making them effectively monopoles with nonAbelian internal degrees of freedom  (i.e., nonAbelian monopoles). 

Such a monopole-vortex connection can be made rigorous mathematically, by use of the exact sequence of homotopy groups, 
\be    \ldots \to   \pi_{2}(G)   \to    \pi_{2}(G/H)   \to      \pi_{1}(H)  \to   \pi_{1} (G) \to \ldots    \label{homotopy}
\ee
applied to the physics situation described above.     As neither of  monopoles and vortices  exists in the full theory   (for $G=SU(N+1)$):  
 \be \pi_2(SU(N+1)) = \pi_1(SU(N+1))= {\mathbbm 1},\ee
the vortex must  end: the endpoints are the monopoles.  
 Each nontrivial element of $\pi_{1}(U(N)) $  -  a vortex -  is associated with a nontrivial element of $\pi_{2}(\frac{SU(N+1)}{U(N)})$
- a monopole.

When  $\pi_1(G)\ne \emptyset$ the situation is a little more subtle:   for instance  $\pi_1(SO(3))= {\mathbf Z}_2$, so in the system
\be     SO(3) \to   U(1) \to \emptyset\;,
\ee 
the minimum winding vortex of the low-energy theory is stable in the full theory.  
The minimum vortex can end at a  ${\mathbf Z}_2$  {\it   magnetic monopole, if introduced in the theory}. 

On the other hand, the doubly-wound vortices cannot exist in the full theory, and that means that 
in the system there must be some object  which absorbs the double magnetic flux carried by the vortex. 
This is how 't Hooft arrived \cite{TP} at the soliton monopole solutions in the first hand.

Note that the mathematics is similar to what happens in the $SU(2)$  Seiberg-Witten theory with the ${\cal N}=1$ perturbation, but physics is different.  The low-energy $U(1)$ in SW model is a magnetic theory, 
thus its breaking generates chromo-electric vortex of minimum winding, which can end only at a source with   chromo-electric  ${\mathbf Z}_2$  charge,  introduced by hand.  But as we noted already,  this is simply a quark in the doublet representation.

In the example (\ref{SUNhierarchy}) the gauge group of the intermediate-scale  theory was $U(N)$, and this fact somewhat obscures (see Table~\ref{GNOduals}) the fact that the vortex flux or 
their source charge (or the sink charge) are classified by the dual group, ${\tilde H}$, not by $H$. 
With other gauge groups, the distinction is clear \cite{KonishiSpanu},\cite{FGK}-\cite{GJK}: see the next section.

\section{NonAbelian vortices \label{sec:NAvort}}  

This brings us to the study of the nonAbelian vortices \cite{Hanany:2003hp}-\cite{Shifman:2004dr}, \cite{Gorsky:2004ad}-\cite{GJK}.  The Abrikosov-Nielsen-Olesen vortices \cite{Abrikosov, NO} arise in an Abelian Higgs model: a scalar $U(1)$ gauge theory in which the potential is such that the scalar field acquires a vacuum expectation value (VEV). In this vacuum a configuration of vortex type, in which 
the scalar field asymptotically (i.e., far from the vortex axis)  winds, i.e.,  acquires a nontrivial phase rotations as its position encircles the vortex axis. According to the scalar quartic coupling with respect to the gauge coupling, the system may produce type I or type II superconductors or BPS  saturated vortices, if $\lambda= g^2/2$ exactly.   

\subsection{Vortices carrying orientational moduli}

NonAbelian vortices arise  in systems such as  (\ref{SUNhierarchy}), with an exact continuous (e.g., $SU(N)$) global symmetry, which is broken by individual vortex solution.  Any soliton solution which  does not respect the global symmetry of the system (e.g.,  a kink solution breaking the translational invariance of the system) develops zeromodes: collective coordinates labelling the solution.  In the case of nonAbelian vortices, one is dealing with the internal, orientational zeromodes, describing the direction of individual solution in the color-flavor diagonal symmetry space. 
 A simple model which possesses such solutions is an $SU(N)\times U(1)$ theory,
 \be   {\cal L}=   \frac{1}{4 g_N^2} (F^a_{\mu \nu})^2 +    \frac{1}{4 e^2} ({\tilde F}_{\mu \nu})^2 +    |{\cal D}_{\mu}  q|^2 -
 \frac{e^2}{2} | q^{\dagger}  q -  c \, {\mathbf 1}|^2 -    \frac{g_N^2}{2} | q^{\dagger} t^a q|^2 \;,
 \ee   
 with an obvious notation. $q$'s  are $N_f=N$ complex scalar "quarks" in the fundamental representation of $SU(N)$. 
 The Bogomolny  completion \cite{Bogomolnyi} gives for static solutions  the coupled linear  differential equations
 \be    ({\cal D}_1 + i \, {\cal D}_2 )\,  q=0\;; 
 \ee
 \be    {\tilde F}_{12}  +   \frac{e^2}{2} (  c\, {\mathbf 1} -   q  q^{\dagger} )=0\;;\qquad  
  {F}_{12}^{(a)}  +   \frac{g_N^2}{2}  q^{\dagger}_i  t^a  q_i =0\;.   
 \ee
 A solution of these equations  (with $q$ written in a mixed color-flavor matrix form) may be taken in the form, 
 \be    q =      U   \,  \left(\begin{array}{cccc}    e^{i \phi}  \phi(r)   &   &   &     \\   & \chi(r)  &   &      \\  &   & \chi(r)   &      \\  &   &   & \ddots      \end{array}\right)  \, U^{\dagger} 
 \label{rotation} \ee
 where $\phi$ is the azimuthal angle around the vortex axis,  $\phi(r)$ and $\chi(r)$ are profile functions. 
 The vortex is oriented,  for $U ={\mathbf 1}$,    in the $(1,1)$ direction in the color-flavor diagonal $SU(N)$ symmetry group.
    The rotation matrix $U$ has the form of the "reducing" matrix \cite{Delduc}, 
 \be   U=   \left(\begin{array}{cc}X^{-1/2}   & - B^{\dagger}  Y^{-1/2} \\BX^{-1/2}   & Y^{-1/2}\end{array}\right)
 \ee
 and 
 \be  B=    \left(\begin{array}{c}b_1 \\b_2 \\\vdots \\b_{N-1}\end{array}\right)
 \ee
 contains the inhomogeneous coordinates of $CP^{N-1}$.
 These constructions have been generalized to other groups \cite{FGK}-\cite{GJK}.

 \subsection{Low-energy effective action: birth of the dual group}
 
 The low-energy effective action for the vortex orientation zeromodes  is obtained by substituting this expression   (\ref{rotation}) in the Lagrangian, taking the $CP^{N-1}$
 coordinates  ${\bf b}$   (the collective coordinates) to be slowly varying functions of $z, t$, and integrating over $(x,y)$ in the plane perpendicular to the vortex axis.   The result is a $2D$  $CP^{N-1}$  sigma model action,
 \be  \const   \int \, dz dt\,  ({\cal D}_{\alpha} n^{c})^{\dagger}  {\cal D}_{\alpha}   n^c\;, \qquad    {\cal D}_{\alpha}  n=  \{ \de_{\alpha} - (n^{\dagger} \de_{\alpha} n) \} n\;, \qquad   n^{\dagger} n=1\;.
 \ee
In a system (\ref{SUNhierarchy}),   the soliton vortex  thus carries fluctuating $CP^{N-1}$ modes;  but in the full theory these vortices must end at the monopoles. As the vacuum of the theory is $SU(N)$ symmetric, and as the monopole-vortex-antimonopole soliton complex {\it as a whole}  breaks it, the monopole and antimonopole act as the source and sink of the propagating $CP^{N-1}$ zeromode fluctuations.  In other words the monopole itself transforms as in ${\underline N}$ of the isometry group. 

{\it  This can be taken as a manifestation of the dual $SU(N)$ gauge group.}  

\begin{figure}
\begin{center}
\includegraphics[width=5in]{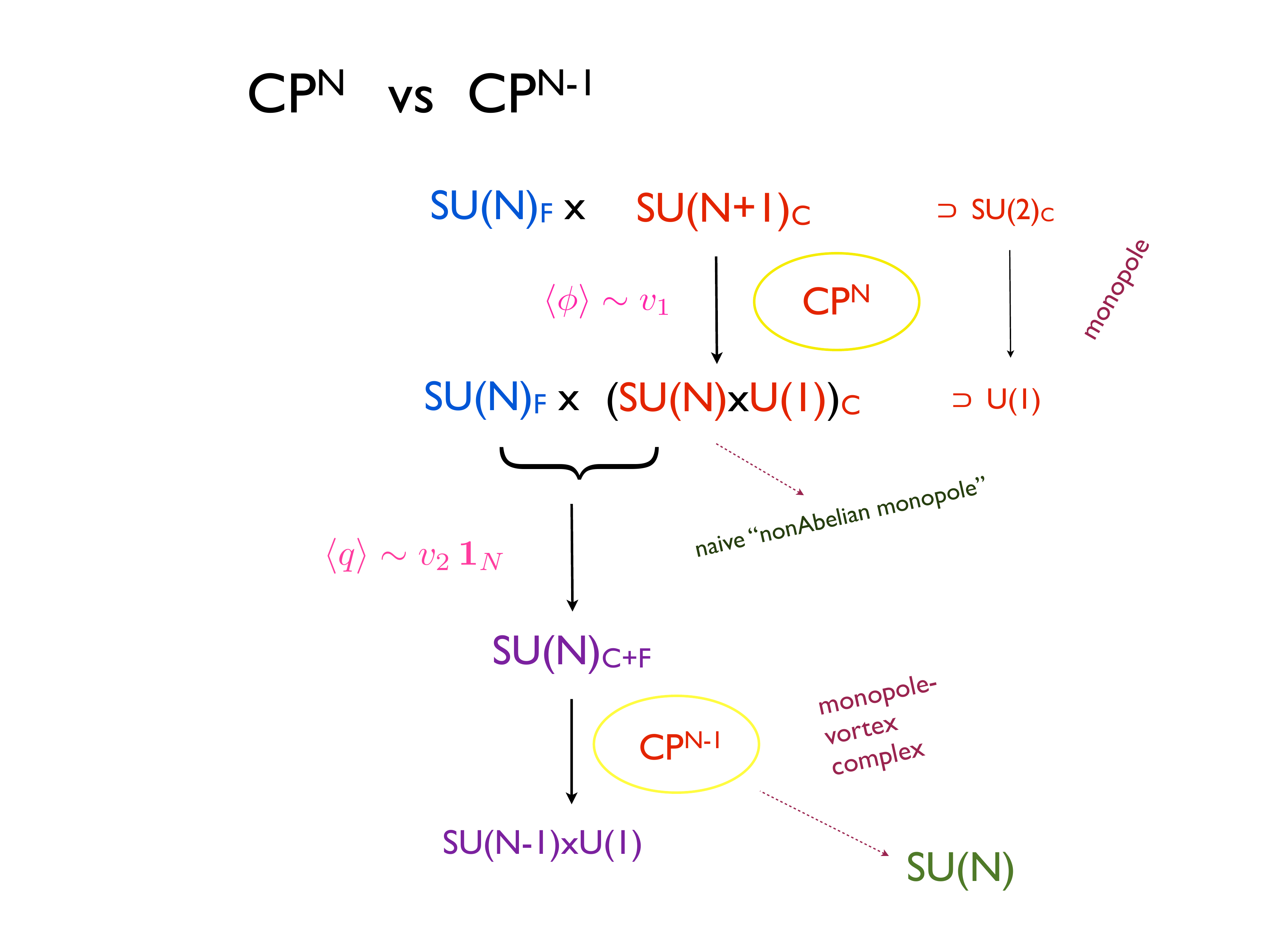}
\caption{ }
\label{twoCPN}
\end{center}
\end{figure}

 The original gauge group being in a Higgs phase, the dual gauge group 
appears in a confinement phase.  The monopole and antimonopole emerge confined, as they should.  
The relation between the na\"{i}ve concept of nonAbelian monopole, and the dual gauge group emerging this way, is illustrated in Fig.~\ref{twoCPN} \footnote{
It is interesting that the flavor $SU(N)$ symmetry group  (in the example discussed above) plays a double role.  On the one hand it is responsible for dressing the monopoles  with flavor quantum numbers (the normalizable fermion $3D$ zeromodes of Jackiw-Rebbi \cite{JR}), affecting their quantum behavior in the infrared.  On the other,  through the color-flavor mixed diagonal symmetry of the vacuum,  
broken by individual vortex-monopole  solution, it gives  rise to the fluctuating $CP^{N-1}$  zeromodes.  Note that the two effects occur at distinct mass scales   (the first at distances $\sim  1 / v_1$, the size of the monopole configuration;  the second involves the vortex worldsheet,  at distances
$\gg 1/ v_2 \gg  1/v_1$). They are distinct effects.}.

As anticipated at the end of the previous section,  the GNO dual of  $U(N)$ is again $U(N)$, and this fact (historically) obscured sometimes the fact the flux of the nonAbelian vortex, hence the charge of their source and sink,  arising in the process $H \to \emptyset$, is classified by the representations of the GNO dual group  ${\tilde H}$.   In fact, for a general gauge group $H$  the distinction is net. For example, in the case of $H= SO(2N)$ theory, with matter in the vectorial representation,  the  minimum vortex fluxes are classified \cite{FGK}-\cite{GJK}  according to  $2^{N-1}$ dimensional spinor representations of positive or negative chirality of  ${\tilde H}= Spin(2N)$. No representations with such dimensions exist in the original $H$ group.

This is perfectly consistent with the idea that the nonAbelian monopoles 
appearing in the system
\be   G \to H  
\ee
 are multiplets of   ${\tilde H}$ \cite{GNO}.   The transformation of the endpoint monopoles in ${\tilde H}$ clearly involves the whole vortex region and 
 is a nonlocal field transformation in terms of the original variables.

How have the difficulties \cite{CDyons}-\cite{DFHK}   associated with the na\"{i}ve concept of nonAbelian monopoles been overcome?   First,    the non-normalizable color $3D$ gauge zeromodes, {\noindent  \it  dressed by flavor, have been converted to normalizable $2D$ zeromodes propagating in the vortex worldsheet.}      As for the topological obstruction,  we note first that a smooth  monopole-vortex configuration is possible only in a gauge in which the scalar field has a fixed orientation asymptotically, i.e., far from the monopole-vortex complex (i.e., does not wind).  The Dirac string singularity attached to the 't Hooft-Polyakov monopole in such a gauge 
lies precisely along the vortex core   \cite{MVComplex3}-\cite{Chatterjee}.  The vanishing of the scalar field there renders the Dirac string physically unobservable, as the field energy density is regular everywhere.

\section{$CP^{N-1}$ sigma model on a finite-width worldsheet  }

\begin{figure}[!ht]
\begin{center}
\mbox{\subfloat{\includegraphics[width=0.49\linewidth]{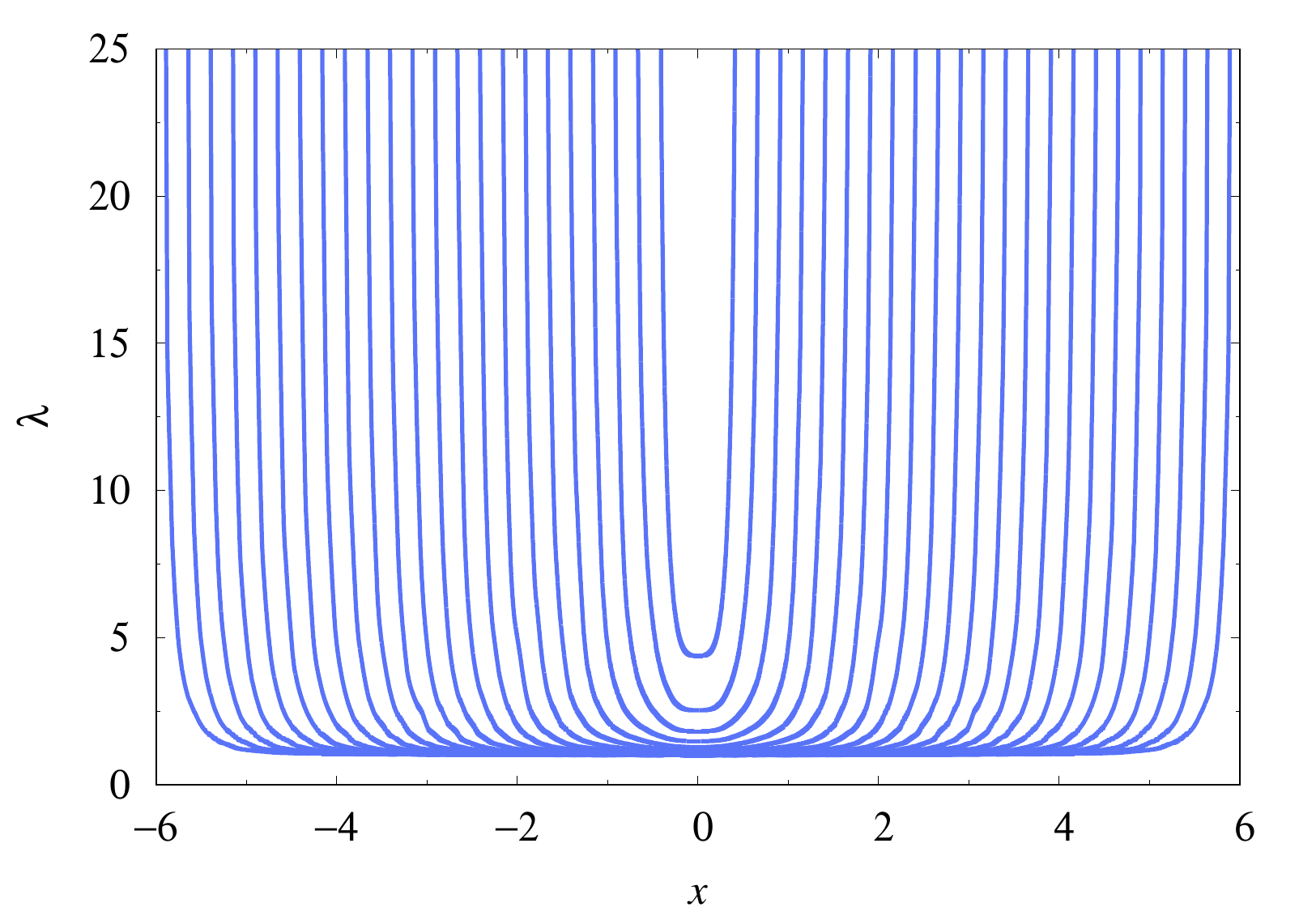}}
\subfloat{\includegraphics[width=0.49\linewidth]{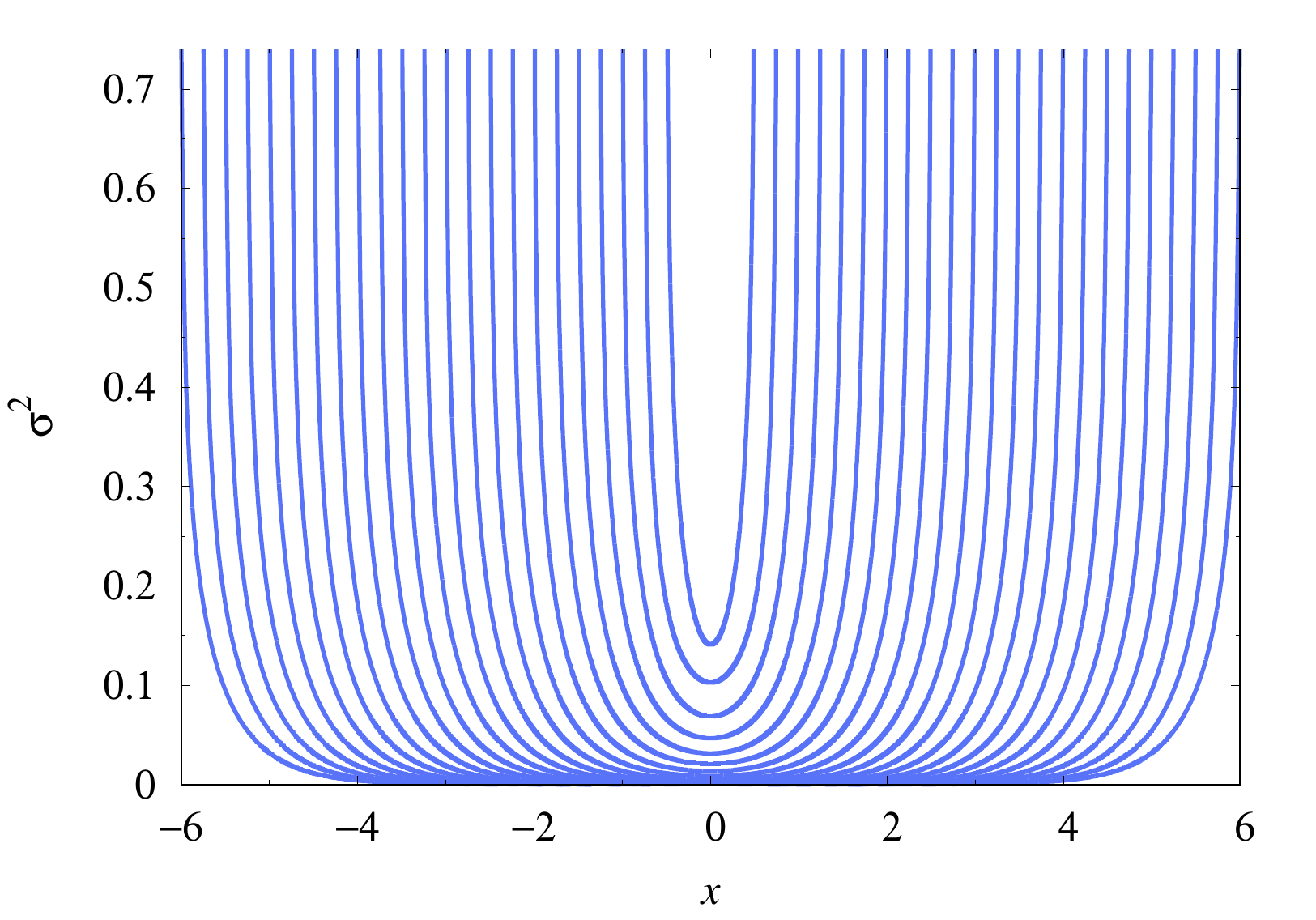}}}\vspace{1em}\\
\caption{The functions  $\lambda(x)$ (left) and   $\sigma^2(x)$ (right)  which are solutions to the gap equation, Eq.~(\ref{gapeqbb}), for 
various values of $L$  ranging  $L=1 \sim 12$.   $\Lambda=1$ in this figure.
The innermost (outermost) curve corresponds to $L=1$ ($L=12$).
}
\label{fig1}
\end{center}
\end{figure}

In order to ensure that the quantum fluctuations of the $CP^{N-1}$  zeromodes do not spoil (i.e., do not break spontaneously) the  $SU(N)$ isometry group, 
we have been led to study the quantum properties of  $2D$ $CP^{N-1}$ sigma model  \cite{DAdda}-\cite{Shifman},    defined on a finite-width worldsheet.  First the nonAbelian vortex-monopole soliton complex have been studied in a semi-classical approximations  \cite{ABEK}-\cite{Chatterjee}.
The quantum physics of this system 
has been investigated systematically only recently \cite{Milekhin1}-\cite{Milekhin2}.  With the boundary condition 
\beq
\hbox{D-D}: \qquad
n_1\!\left(-\tfrac{L}{2}\right)=n_1\!\left(\tfrac{L}{2}\right) =
\sqrt{r}\;,   \qquad
n_{i}\!\left(-\tfrac{L}{2}\right)=  n_{i}\!\left(\tfrac{L}{2}\right) =0\;,  \quad i>1\;.  \label{DDbc}
\eeq
the effective action has the form, 
\beq  
{S}_{{\rm eff}} = \int d^2 x \left( (N-1) \, {\rm tr}\, {\rm log} (- D_{\mu}D^{\mu}  + \lambda) + (D_{\mu} \nc)^*D^{\mu} \nc -  \lambda (|\nc|^2  - r)  \right)\;.\label{from}
\eeq
which leads to the generalized gap equations, 
\beq
\frac{N}{2} \, \sum_n\frac{f_n(x)^2}{\omega_n} e^{-\epsilon \omega_n}  + \nc(x)^2 - r_{\epsilon} = 0  \,,\qquad  \partial_x^2 \nc(x) - \lambda(x) \nc(x) = 0  \,,  
\label{gapeqbb}
\eea
where
\begin{eqnarray}
r_\epsilon \equiv 
\frac{N}{2\pi}\left(\log\left(\frac{2}{\Lambda  \epsilon}\right)
-\gamma \right)\;,  
 \label{uvbeta}
\end{eqnarray}
The system has been studied both analytically and numerically.  We find \cite{BKO,BBGKO}  a unique solution of the coupled gap equations for any $L$, which in the limit of $L \to \infty$ smoothly goes over to the well-known $2D$  $CP^{N-1}$ sigma model vacuum \cite{DAdda}-\cite{Shifman},  with mass gap and without  spontaneous breaking of the isometry $SU(N)$ group  (i.e. no $n_1=\sigma$ condensate. 
Some numerical results for the gap function $\lambda(x)$ and the classical field $\sigma(x)$ are shown in Fig.~\ref{fig1}.

\begin{figure}[!ht]
\begin{center}
\mbox{\subfloat{\includegraphics[width=0.49\linewidth]{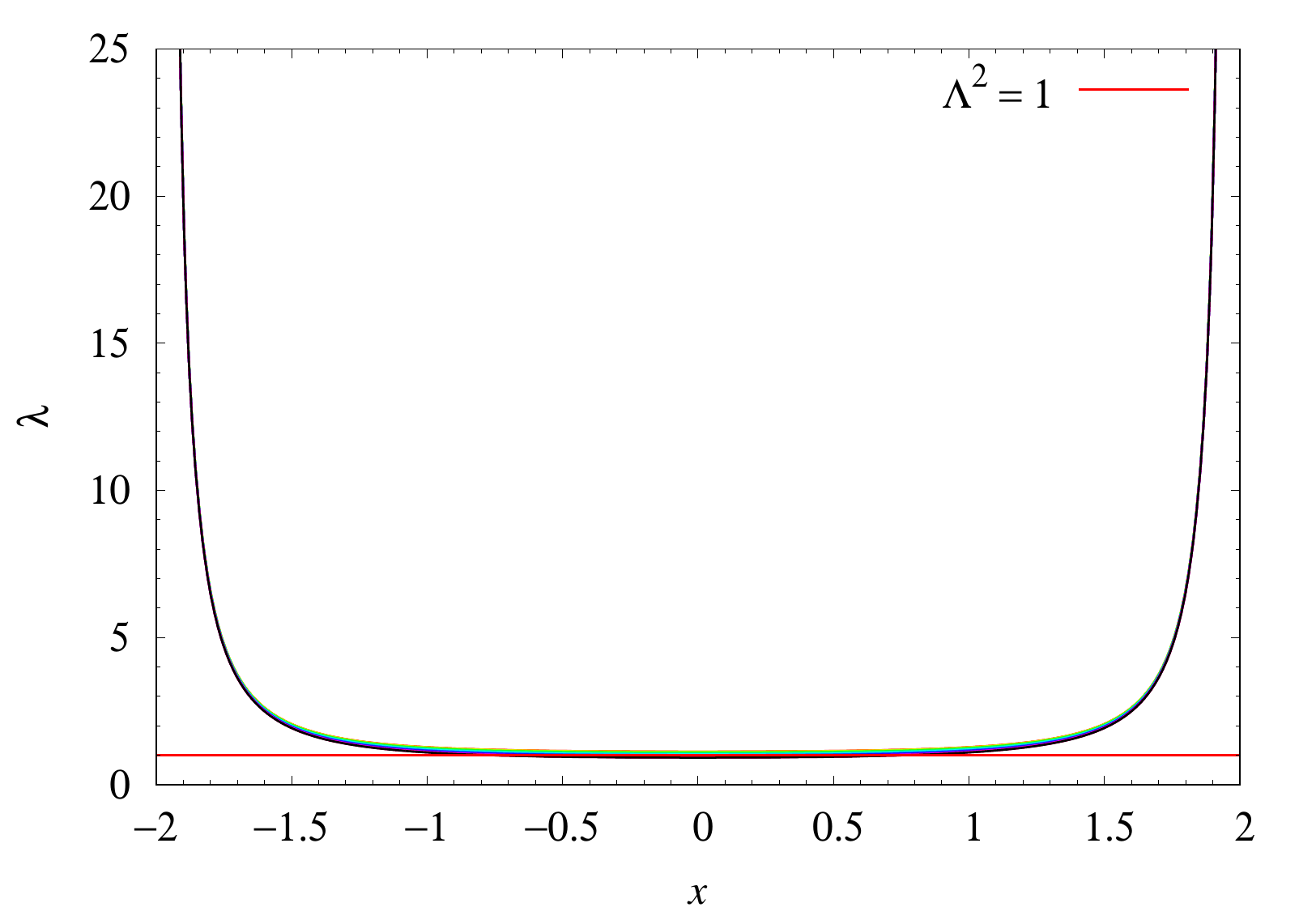}}
\subfloat{\includegraphics[width=0.49\linewidth]{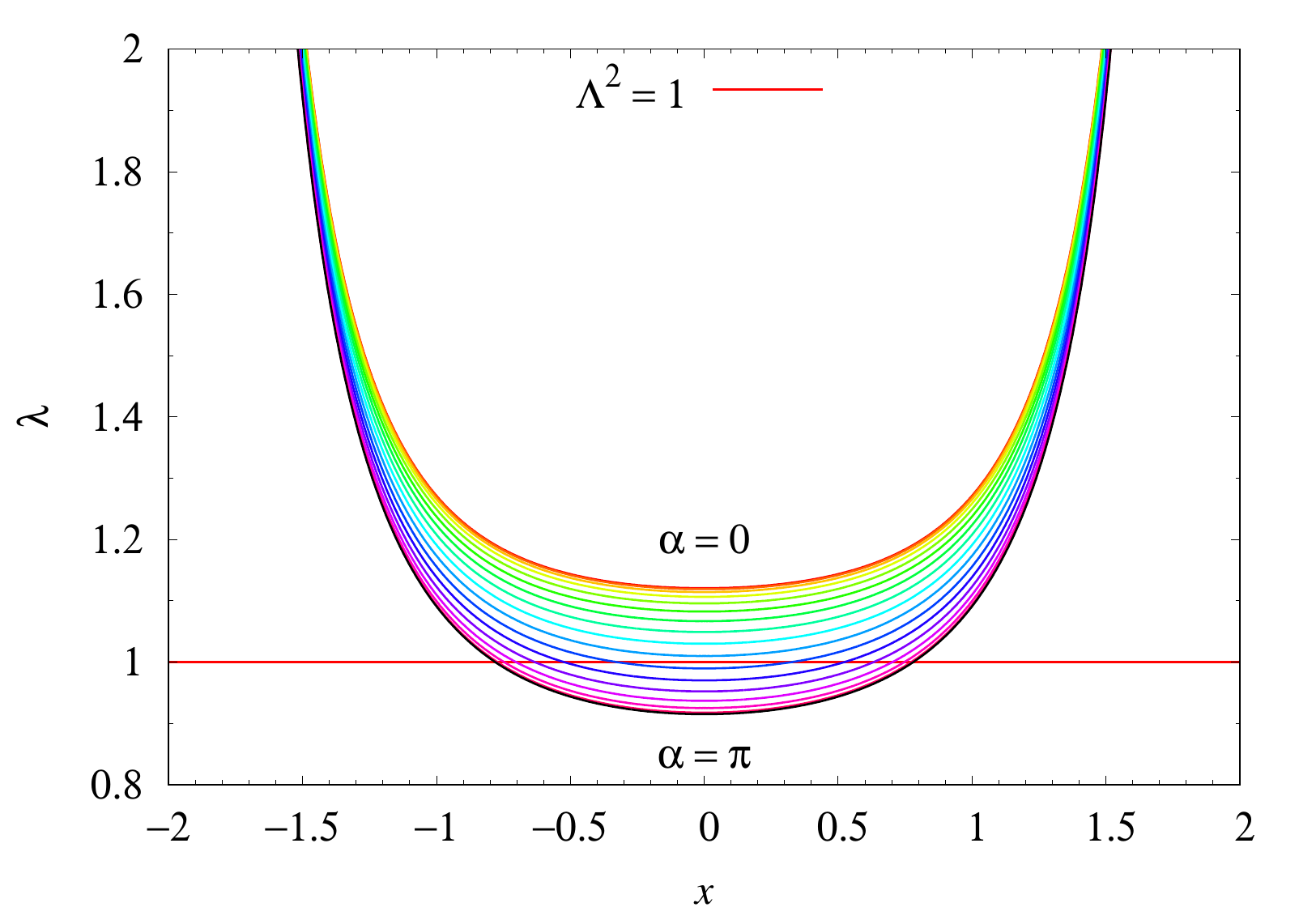}}}
\caption{\small The gap function  $\lambda(x)$ which solves the gap
  equation is plotted for various values of
  $\alpha$ for  $L=4$ and  $\Lambda=1$.
  $\alpha=0,\tfrac{\pi}{16},\tfrac{2 \pi}{16}, \ldots, \pi$ from the
  top curve to the bottom.
  On the right are the same curves zoomed in,  on the vertical.
}
\label{fig:lambda}
\end{center}
\end{figure}

\begin{figure}[!ht]
\begin{center}
\mbox{\subfloat{\includegraphics[width=0.49\linewidth]{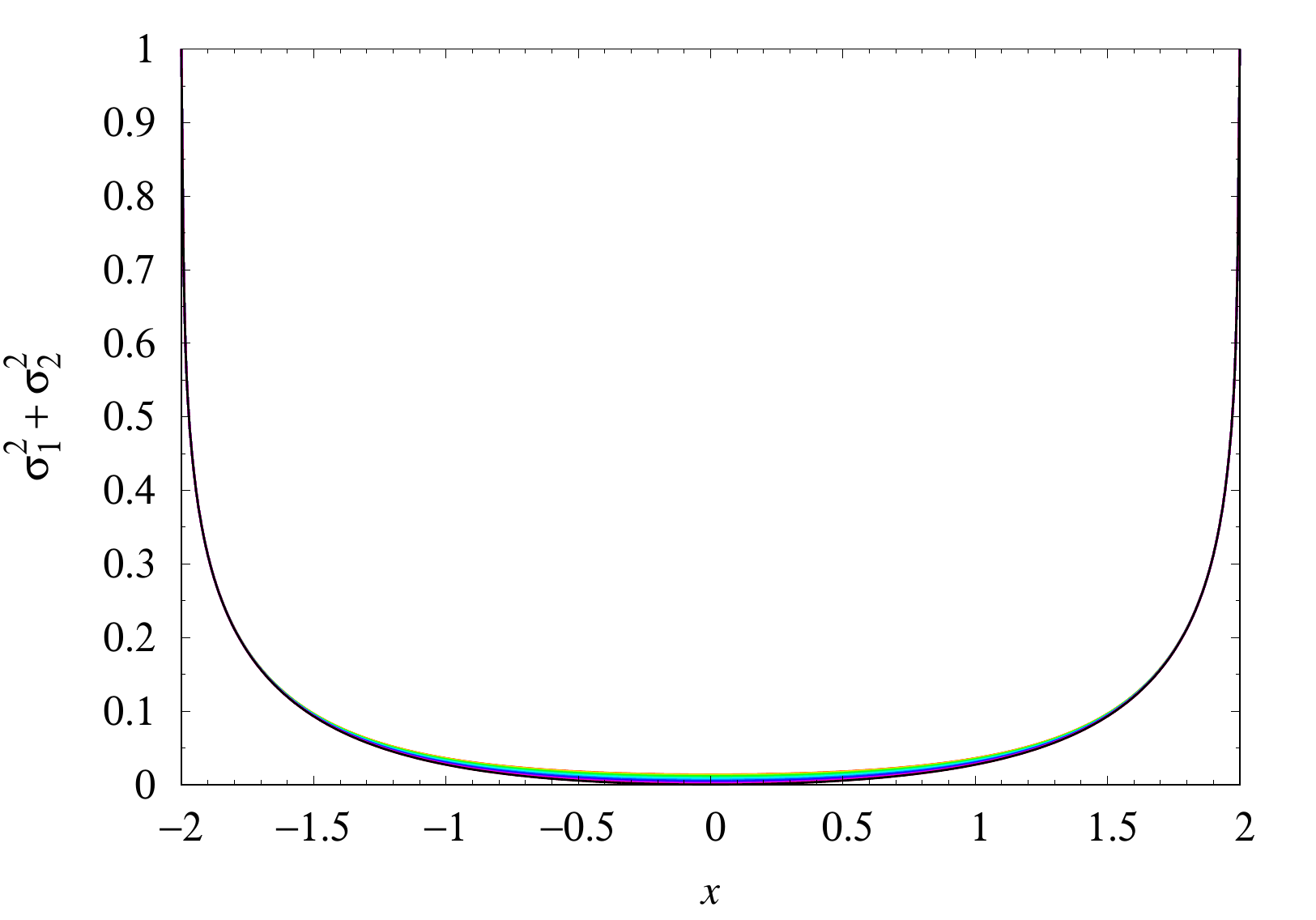}}
\subfloat{\includegraphics[width=0.49\linewidth]{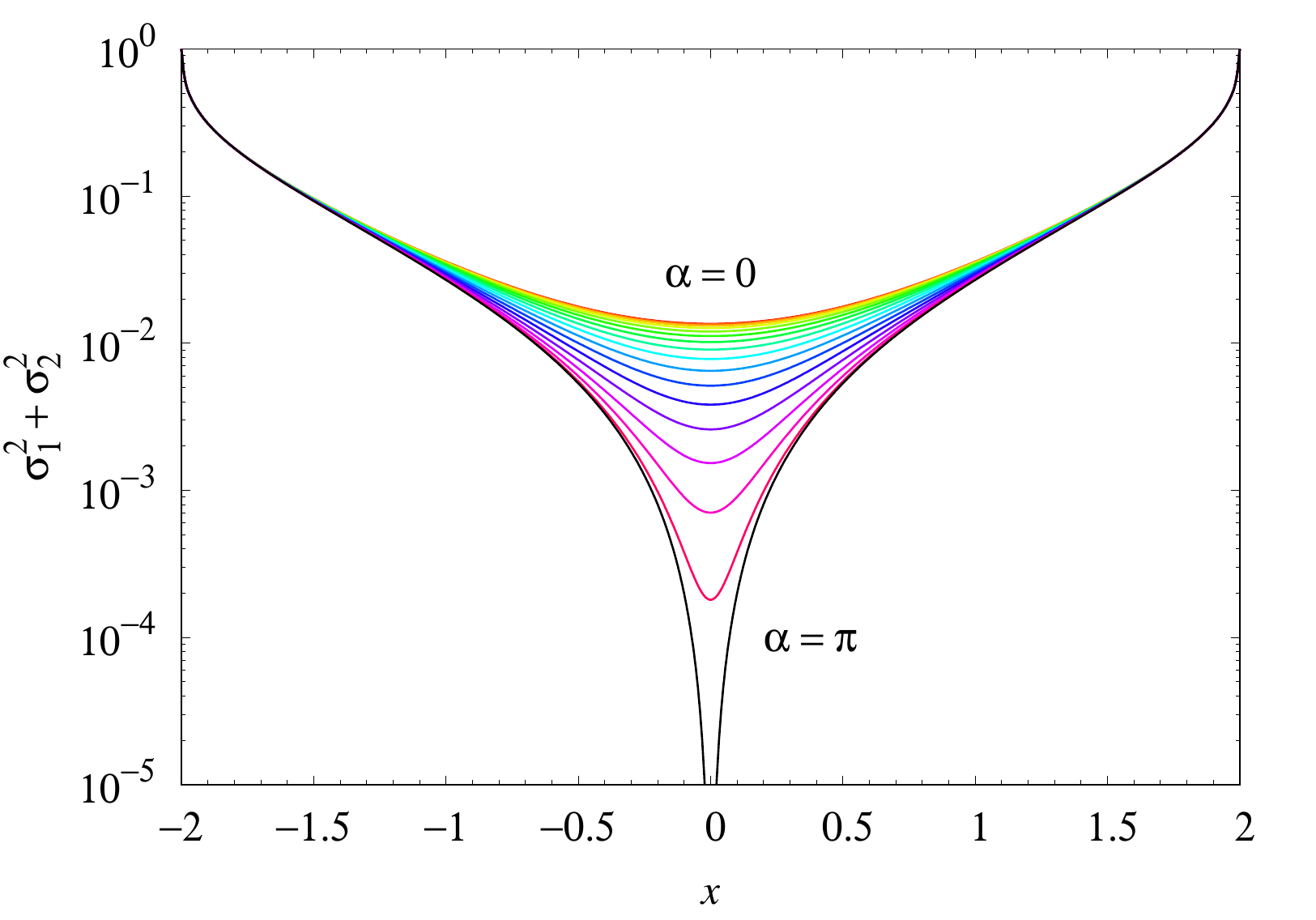}}}
\caption{\small The function  $\sigma_1(x)^2+\sigma_2(x)^2$  is shown
  in a normal plot (left) and on a logarithmic scale (right), for
  various values of $\alpha$ for  $L=4$ and  $\Lambda=1$.  The values
  of $\alpha$ are
  $0, \tfrac{\pi}{16}, \tfrac{2 \pi}{16},\ldots, \pi$ from the top to
  the bottom curves.
  }
\label{fig:ssum}
\end{center}
\end{figure}

The cases with the most general Dirichlet boundary condition, where the orientations in $CP^{N-1}$ at the two boundaries are different and generic, 
have also been studied recently \cite{BGKO}.   The boundary conditions 
look like
\beq   
\left(\begin{array}{c}n_1\!\left(\tfrac{L}{2}\right) \\n_2\!\left(\tfrac{L}{2}\right)\end{array}\right) =  \left(\begin{array}{c} 1 \\0\end{array}\right) \sqrt {r_{\epsilon}} \;;
\label{bc1} 
\eeq 
\beq 
  \left(\begin{array}{c}n_1\!\left(-\tfrac{L}{2}\right) \\n_2\!\left(-\tfrac{L}{2}\right)\end{array}\right) = \left(\begin{array}{cc}e^{i \gamma} \cos \alpha  & e^{i \beta} \sin \alpha \\- e^{-i\beta}\sin \alpha & e^{-i \gamma} \cos \alpha \end{array}\right) \left(\begin{array}{c} {\sqrt{r_{\epsilon}}}  \\0\end{array}\right)  \sim 
   \left(\begin{array}{c} \cos \alpha  \\   \sin \alpha\end{array}\right)  \sqrt {r_{\epsilon}}\;;    \label{bc2}
\eeq
\beq   
n_{i}\!\left(-\tfrac{L}{2}\right)=  n_{i}\!\left(\tfrac{L}{2}\right)=0\;,  \qquad i>2\;.    \label{bc2bis} 
\eeq
where the gap function $\lambda(x)$ and two classical components are determined from the gap equation, 
\beq
\frac{N}{2} \, \sum_n\frac{f_n(x)^2}{\omega_n} e^{-\epsilon \omega_n}  + |\sigma_1(x)|^2 +   |\sigma_2(x)|^2   - r_{\epsilon} = 0  \,,  \label{gapeqbbGen1} \eeq
\beq
 \partial_x^2 \sigma_1 - \lambda(x) \sigma_1 = 0  \,,  \qquad   \partial_x^2 \sigma_2 - \lambda(x) \sigma_2 = 0\;.
\label{gapeqbbGen2}
\eea
Some  numerical results for the mass gap function (for $L=4$, $\Lambda=1$) are shown in Fig.~\ref{fig:lambda},
and for the classical components in  Fig.~\ref{fig:ssum}.

The basic conclusion from these studies is that the global, isometry $SU(N)$ symmetry of the $CP^{N-1}$ model is not spontaneously broken at finite $L$,  as 
in the $L=\infty$ case \cite{DAdda}-\cite{Shifman}.
The concept that the boundary  magnetic monopoles behave according to  ${\underline N}$ of the new, dual $SU(N)$ group, is therefore valid quantum mechanically.

\section{Strongly-coupled nonAbelian monopoles in action}

\subsection{Confining vacua near strongly-coupled IRFP}

Having established the notion of quantum-mechanical nonAbelian monopoles, we now go back to ${\cal N}=2$ supersymmetric QCD
 and try to study
them "in action".
Do ${\cal N}=2$ SQCD teach us something useful on the confinement vacuum in which strongly-coupled  nonAbelian monopoles play  the
central role?   

\begin{figure}
\begin{center}
\includegraphics[width=5in]{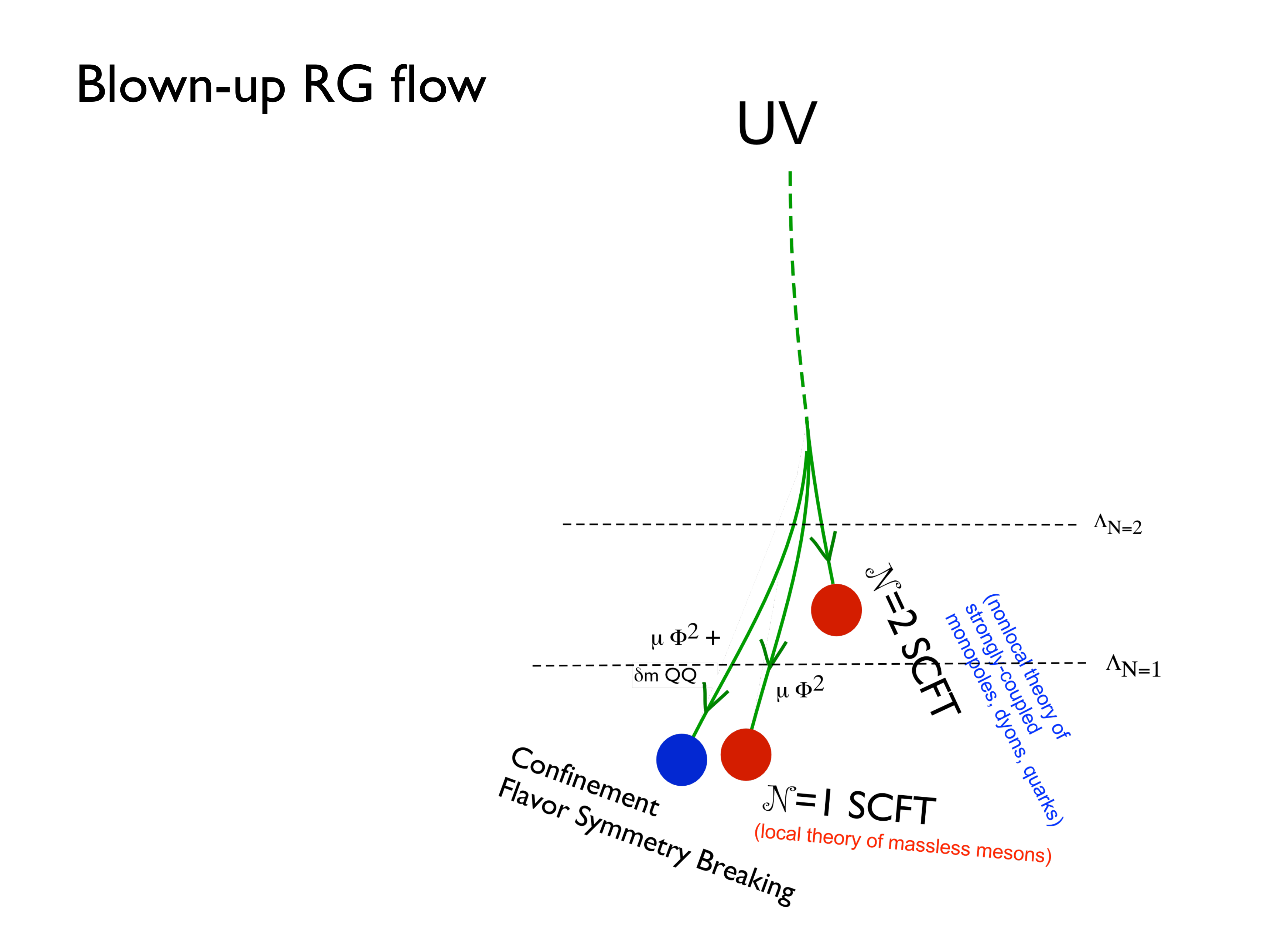}
\caption{ }
\label{BlownUp1}
\end{center}
\end{figure}

A recent interesting observation \cite{Bolognesi:2015wta} is that the most singular ("Argyres-Douglas" \cite{AD,Argyres:1995xn}) superconformal theories (SCFT) in ${\cal N}=2$, $SU(2)$ theory with $N_f=1,2,3$ and $SU(N)$ theory with $N_f$ flavors,  flow down, under an ${\cal N}=1$ perturbation, 
\be    \mu\, \Phi^2 |_F =   \mu \psi \psi + \ldots   \label{N=1def}
\ee
towards  an infrared-fixed-point theory, {\it described by massless mesons $M$ in the adjoint representation of $G_F$.  }
Further relevant deformations (shift of bare masses from the critical value) leads to confinement and   flavor symmetry breaking.  The meson fields $M$ 
(part of them) are now the massless Nambu-Goldstone bosons.   A blown-up picture of these RG flows is shown in Fig.~\ref{BlownUp1}.

What is it that is so remarkable about it?  There are at least several issues: 
\begin{enumerate}
  \item    ${\cal N}=2$ SCFT, which is the destiny of the system  under the renormalization-group (RG) flow,  without ${\cal N}=1$ deformation,  is a complicated nonlocal theory of strongly-coupled
  massless nonAbelian monopoles, dyons and quarks \footnote{It is for this reason that the study reported here became possible only after the work by Argyres and Seiberg, Gaiotto and others \cite{Argyres:2007cn}-\cite{tienne}.  
  };   
  \item   ${\cal N}=1$ SCFT  towards which the RG flow is steered by the deformation (\ref{N=1def}),  is instead a theory of weakly-coupled local, infrared-free  theory of gauge-inavriant mesons in the adjoint representation of the original flavor symmetry. 
  \item In the nearby  ${\cal N}=1$ confining vacuum, the massless mesons $M$ now become NG bosons of symmetry breaking. It appears as if the  
${\cal N}=1$ SCFT  were "preparing"  for the necessity of giving rise to the right NG bosons, in case some further symmetry breaking arises in the system.
\item  The situation seems to be somewhat analogous \footnote{Analogy, as beauty, lies in the eyes of those who see it, according to the author.
The point could be that massless mesons in the adjoint representation (related to the generators of the symmetry group) are a signal of a flavor symmetry breaking.  
} to what happens in the real-world  (${\cal N}=0$) QCD. 

\end{enumerate}

\subsection{Softly broken ${\cal N}=2$  SQCD}

Now how does all this come about?  The basic tool is the Seiberg-Witten curves for the ${\cal N}=2$ theories.  As discovered by Argyres-Douglas (AD)  and others around '95 \cite{AD}-\cite{Eguchi},  SCFT  may appear in many gauge theories,  when the flavor group, the mass parameters, and the vacuum moduli points,  etc. are appropriately tuned.
The interest in exploring SCFT in  $4D$  ${\cal N}=2$ gauge theories has been more recently revived after the work by Argyres-Seiberg, Shapere-Tachikawa, Gaiotto, and others, and now it has become a research field of rare beauty and depth \cite{Argyres:2007cn}-\cite{tienne}. 
 
Especially, we are interested in the most singular AD SCFT in  ${\cal N}=2$  SQCD  with $SU(N)$ gauge group and $N_f$ flavors of  quark multiplets. 
One particularly powerful tool is the trace (conformal) anomaly:  in any classically conformal theory the quantum effects give for the trace of enegy-momentum tensor
\be    \brc T^{\mu}_{\mu} \ckt =  \frac{1}{16 \pi^2}  \left[   c\, (R_{\mu \nu \rho \sigma}^2 - 2 R_{\mu \nu}^2 +  \frac{R^2}{3} ) - a\, 
(R_{\mu \nu \rho \sigma}^2 -  4  R_{\mu \nu}^2 +  R^2 ) 
\right] 
\ee
where the $c$ and $a$  coefficients 
appear in front of the Weyl tensor square and  of the Euler density. They depend on the massless particles present in the system:   the (gauge) vectors,  fermions and scalars.  
For any ${\cal N}=1$ supersymmetric  theory, a powerful result \cite{anselmi} is that the $a$ and $c$ coefficients are related to the trace (sum over Weyl fermions) 
$\Tr \,R^3$ and $\Tr \,R$   as
\be  a=  \frac{3}{32} ( 3 \,   \Tr \,R^3 - \Tr \,R )\;; \qquad c = \frac{1}{32} ( 9  \, \Tr \,R^3 -  5 \, \Tr \,R )  \;,
\ee
where $R$ is the $R$ charge present in the ${\cal N}=1$ supersymmetry algebra.   
An ${\cal N}=2$  supersymmetric model, instead, has  a global 
\be   SU_R(2) \times U_R(1) 
\ee
symmetry.   Denoting the $R$ charge of the ${\cal N}=2$ theory as $R_{{\cal N}=2}$,  now there are relations
\be    \Tr \, R_{{\cal N}=2}^3 =  \Tr \, R_{{\cal N}=2}=  48 \, (a-c)\;, \qquad  
\Tr \, R_{{\cal N}=2}  I_a I_b =  \delta_{ab}   ( 4 a - 2 c)\,.  
\ee 
where $I_a$ denote the $SU_R(2)$ "isospin". 

Flowing down from  near ${\cal N}=2$ SCFT  towards a new, ${\cal N}=1$ SCFT  under the deformation 
\be   \mu \Phi^2  |_{\theta^2} 
\ee
can be studied by using the results by Bonelli et.al \cite{noi} and by Giacomelli \cite{SGnuovo}, who complemented the ${\cal N}=2$ curve with the factorization condition  (i.e., condition that the theory survives the ${\cal N}=1$ perturbation), getting the ${\cal N}=1$ curves.  Requiring that the surviving  ${\cal N}=1$ vacuum 
is conformal, leads to the relation between the $R$ charges 
\be    \{ R_{{\cal N}=2},  I_3\} \longleftrightarrow     R_{{\cal N}=1} \;.  
\ee
For example, these relations read explicitly \cite{Bolognesi:2015wta}
\be     R_{{\cal N}=1}  =   \frac{5}{6}  R_{{\cal N}=2} +  \frac{1}{3} I_3\;, \qquad   {\rm for} \quad SU(2), \quad N_f=1\;;
\ee
\be     R_{{\cal N}=1}  =   \frac{2}{3}  R_{{\cal N}=2} +  \frac{2}{3} I_3\;, \qquad   {\rm for} \quad SU(N), \quad N_f=N-1\;.
\ee
These relations allow to derive, from the known traces involving $\{ R_{{\cal N}=2},  I_3\} $  of a given ${\cal N}=2$  SCFT \cite{TS}-\cite{iocecotti},   the values of  
\be    \Tr  \, R_{{\cal N}=1}^3\;, \qquad   \Tr  \, R_{{\cal N}=1}\;,
\ee
(the trace means the sum over the massless Weyl fermions present);  these, on the other hand, allow us to find out the $a$ and $c$ coefficients of the ${\cal N}=1$  infrared-fixed-point theory.  For instance,  one gets 
\be    a_{IR} = \frac{1}{48}\;, \qquad c= \frac{1}{24}\;,     
\ee
for $SU(2)$, $N_f=1$,
\be    a_{IR} = \frac{1}{6}\;, \qquad c= \frac{1}{3}\;,  \label{IR1} 
\ee
for $SU(2)$, $N_f=3$, and so on.  For  the AD point of the $SU(N)$ theory with $N_f=2N-1$,  the relation between ${\cal N}=2$ $R$ charges  and   ${\cal N}=1$ $R$ charge is given by
\be    R_{{\cal N}=1}=   \frac{2}{3}   (  R_{{\cal N}=2} + I_3 )\;.   \label{IR2} 
\ee
By saturating the 't Hooft anomaly matching conditions for 
\be      {\Tr}  \, R_{{\cal N}=1}^3\;, \qquad  \Tr  \, R_{{\cal N}=1}\;, \qquad    \Tr  \, R_{{\cal N}=1}  G_f^2\;, 
\ee
one finds that the $a$ and $c$ coefficients of the infrared-fixed-point theory are given by
\be    a_{IR}=  \frac{ (2N-1)^2 -1}{48}\;; \qquad   c_{IR}=  \frac{ (2N-1)^2 -1}{24}\;. \label{IR3} 
\ee
Note that  in all cases, (\ref{IR1}), (\ref{IR2}), (\ref{IR3}),  the conformal anomaly coefficients are consistent with the hypothesis that 
the  ${\cal N}=1$  infrared-fixed-point theory is a theory of  (infrared-)free massless mesons  $M$  described by  chiral multiplets  in the adjoint representation of the flavor group
$G_f$. 
It is interesting the 't Hooft anomaly matching condition for  $  {\Tr}  \, G_f^3  $  is trivially satisfied as the UV (${\cal N}=2$ SCFT)  is an ${\cal N}=2$ vectorlike theory and the low-energy chiral multiplets in ${\cal N}=1$ SCFT  are in the adjoint representation.

\begin{figure}
\begin{center}
\includegraphics[width=5in]{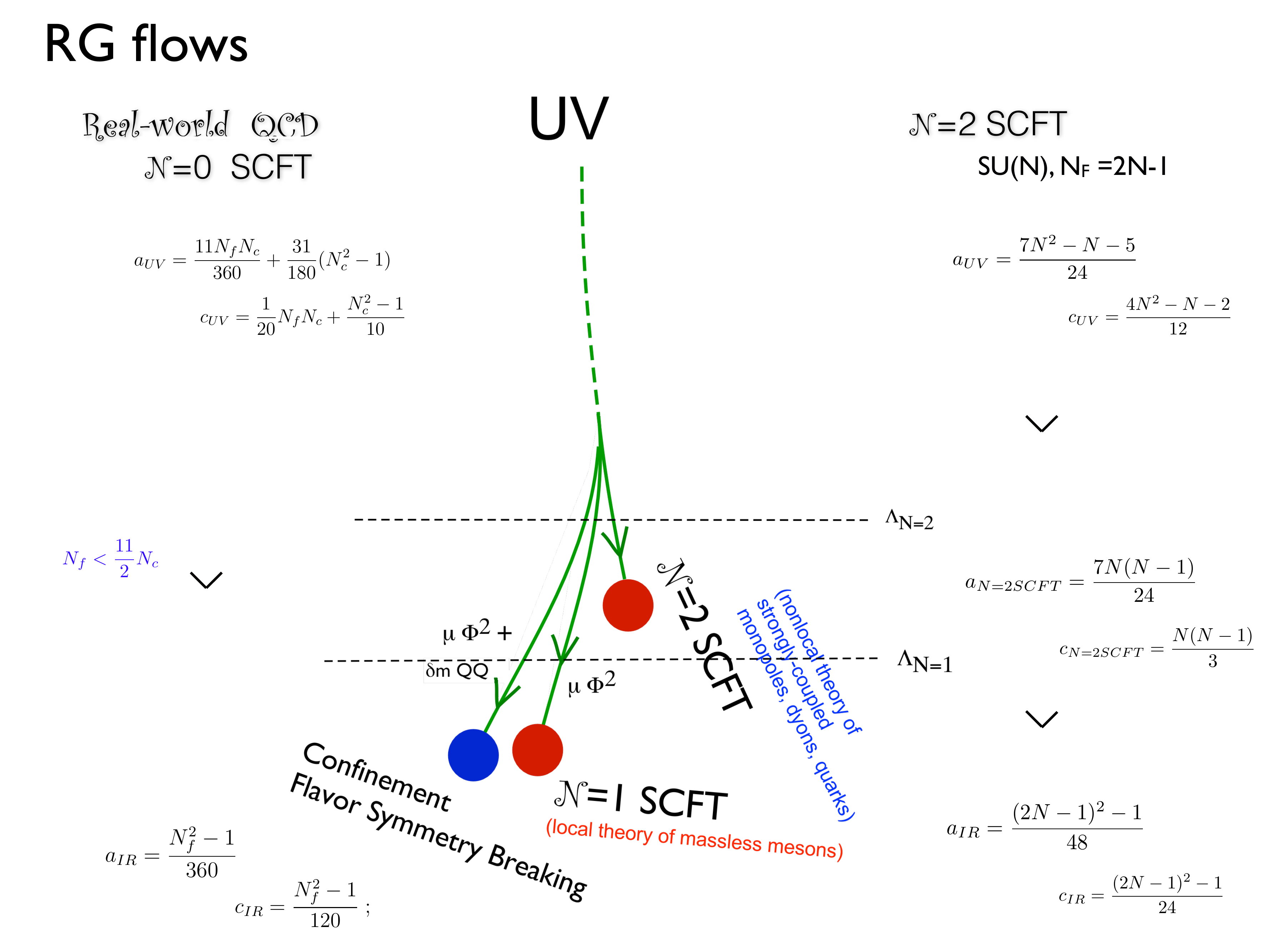}
\caption{ }
\label{BlownUp2}
\end{center}
\end{figure}

All in all, let us illustrate the blown-up RG flow diagram again, Fig.~\ref{BlownUp2}, with the conformal coefficients added.   We observe that indeed, the $a$ theorem (that $a$ decreases as one goes towards the infrared, showing the loss of information) is satisfied in all cases (the actual values are shown only for $SU(N), N_f=2N-1$ case.)

Finally, a further small perturbation which shifts the common bare quark masses slightly off the critical value 
 (needed for the system to go to an SCFT IR fixed point),  induces the condensation of these meson fields, and the system goes into confinement phase, with global symmetry breaking.

The flow of the  conformal anomaly coefficients for the ordinary QCD is also shown for  comparison. 

\section{Confinement and XSB in "chiral" QCD \label{psichieta} }

A final topics I would like to touch upon is the confinement and chiral symmetry breaking in a "chiral" variety of QCD, recently 
 discussed by Bolognesi, Shifman and myself \cite{BKS}.  In particular, our interest is to understand the dynamics of the $\psi\chi\eta$ model, namely an $SU(N)$
 theory with matter (left-handed)  Weyl fermions,
 \be     \psi^{\{ij\}}\;, \qquad   \chi_{[ij]}\;,  \qquad  \eta_i^A\;, \qquad (A=1,2,\ldots 8)\;.
 \ee
 in the representations
 \be  \!\!\!\!  \yng(2)\;, \qquad  {\bar {\yng(1,1)}}\;, \qquad   {\bar  {\yng(1)}}\;,
 \ee
 respectively.  This theory  has a global symmetry,
 \be     G_f=  SU(8) \times U_1(1)\times U_2(1) \times {\mathbf Z}_{N^*}\;,
 \ee 
 where $U_{1,2}$ are anomaly-free combinations of the three $U(1)$'s associate with the $\psi$, $\chi$ and $\eta$ fields,  
 \be    U_1(1): \qquad \qquad   \psi \to e^{i \tfrac{\alpha}{N+2}} \psi\;, \qquad   \eta  \to  e^{- i \tfrac{\alpha}{8} } \eta\;;    \label{U11}
 \ee
 \be    U_2(1): \qquad \qquad   \psi \to e^{i \tfrac{\beta}{N+2}} \psi\;, \qquad   \chi   \to  e^{- i \tfrac{\beta}{N-2} }  \chi\; \label{U12}
 \ee
 and 
 \be       {\mathbf Z}_{N^*}\;:   \quad \quad  N^*=  GCD  \{ N+2, N-2, 8\}\;,
 \ee
 is a combination of anomaly-free discrete subgroups of three $U(1)$'s  which do not belong to $U_1(1) \times U_2(1)$.  
This model is  asymptotically free, and the interactions become strong in the infrared \footnote{There is a long history of studies of chiral varieties of QCD; see \cite{Raby}-\cite{AS}. }.
No gauge-invariant bifermion scalar composites can be formed, and one may wonder what the fate of the global $SU(8)$ symmetry 
could be.  The simplest scenario could be that the system confines, but no fermion condensates are formed. The chiral $SU(8)$ symmetry 
would remain intact, and it seems to be a hopeless task to find a set of gauge-invariant massless baryons which can saturate the 't Hooft anomaly. 

It is possible that four-fermion (gauge-invariant) condensates are formed, so that (for instance) the flavor $G_f$ symmetry is completely broken. 
It is however hard to realize this scenario when it comes to details \cite{BKS}.

We instead assume that  two types of fermion bilinear condensates,
\be
 \langle   \phi^{i A}  \rangle \sim  \langle  \psi^{ij}  \eta_j^A   \rangle \,,\quad A= 1,2, \ldots, 8\,,
\ee
and 
\be  \,\,\,  \langle   {\tilde \phi}^{i}_j \rangle = \langle  \psi^{ik} \chi_{kj}   \rangle\;,\label{adjcond}
   \ee
 play the crucial role in the dynamics.  The first one is in the fundamental representation of the gauge group, while the second is in the adjoint.  This leads to two possible dynamical scenarios in this model.  
 
 \subsection{Partial color-flavor locking and dynamical Abelianization  (for $N \ge 12$)}
 Let us assume the condensates of the form, 
 \beq
 \brc  \phi^{i A}   \ckt  = \Lambda^3 \left(\begin{array}{c} c  {\mathbf 1}_{8}  \\ \hline
  \\
{\mathbf 0}_{N-8,8}\\
\\
\end{array}\right)\;, \qquad 
 \langle  \tilde{\phi}^{i}_j \rangle  = \Lambda^3   \left(\begin{array}{c|ccc|c}
 a \,  {\mathbf 1}_{8}   &  &  & &  \\ \hline  &  d_1  &  & & \\ &   & \ddots &  & \\ &  &  & d_{N-12}&
\\ \hline & &  &&   b  \,  {\mathbf 1}_{4}   \end{array}\right)\;,   \label{conden}
\eeq
where 
\be    8 a + \sum_{i=1}^{N-12}  d_i + 4  b =0\;, 
\qquad     a, d_i, b \sim    O(1)\;. 
\ee
These will break the symmetry as
\beq
SU(N)_{\rm c} \times SU(8)_{\rm f}   \times U(1)^2  \to   SU(8)_{\rm cf} \times  U(1)^{N-11} \times SU(4)_{\rm c}\,.
\label{scenario1}
\eeq
Now the $SU(8)$ 't Hooft anomaly triangles are saturated as follows. In the UV it is simply given by the $\eta$ particles: it is  $N$. 
In the IR,  the baryons
\beq
      B^{\{AB \}}=\psi^{\{ij\}} \eta_j^A \eta_i^B\Big|_{A,B \,\, {\rm symm} }  \sim     \langle \phi^{i A} \rangle \eta_i^B\Big|_{A,B \,\, {\rm symm} }    \eeq
transforming in the symmetric representation of 
$SU(8)_{\rm cf}$,  gives   $8+4=12$; and  the weakly coupled 
\beq      
{\tilde B}^A_j=\psi^{ik} \chi_{kj} \eta_i^A   \sim   \langle {\tilde \phi}^{i}_j\rangle   \eta_i^A   \, , \qquad j=9,10,\ldots, N-4
\label{baryonp}
\eeq
contribute   $ N-12$,  so that  their total equal to  the UV value, $N$. 

Note that the adjoint condensate  (\ref{adjcond})  breaks spontaneously both nonanomalous $U(1)$ symmetries  (\ref{U11}), (\ref{U12}). This is crucial. There are two NG bosons.  

Actually the color-flavor locked $SU(8)$ might be broken to smaller (CF locked) group, 
\beq
SU(N)_{\rm c} \times SU(8)_{\rm f}   \times U(1)^2  \to   \prod_i SU(A_i)_{\rm cf} \times  \prod U(1)_{\rm cf}  \times    U(1)^{N-11
} \times SU(4)_{\rm c}\,.
\label{scenario1bis}
\eeq
but one can show that the 't Hooft anomaly matching condition for  any  triangles involving $\prod_i SU(A_i)_{\rm cf} \times  \prod U(1)_{\rm cf} $
are all saturated by the same massless fermions $B^{\{AB \}}$ and  ${\tilde B}^A_j$.

 \subsection{Full Abelianization} 
 
 Another possibility, which seems to be the only viable one for $N < 12$,  is that the adjoint condensate
 takes the form, 
 \be    \brc {\tilde \phi}^i_j \ckt =   \brc \psi^i \chi_{kj} \ckt=\Lambda^3  \,  \left(\begin{array}{ccc}d_1 &  &  \\ & \ddots &  \\ &  & d_N\end{array}\right)\;;
 \ee
 the symmetry is broken as 
   \beq     
  SU(N)_{\rm c} \times SU(8)_{\rm f} \times U(1)^2 \to U(1)^{N-1} \times  SU(8)_{\rm f}\,,
  \label{scenario3}
  \eeq
  and the weakly coupled 
\be  {\tilde B}^A_j  \sim \eta_j^A\;, 
\ee
trivially saturate the anomaly matching condition.  

To summarize:   in the $\psi-\chi-\eta$  model   either 
partial color-flavor locking and dynamical Abelianization, or  full dynamical Abelianization,  seems to occur. 
Planar equivalence to Supersymmetric YM, conjectured earlier \cite{AS},  does not hold.   Dynamical Abelianization which is ubiquitous in ${\cal N}=2$ 
supersymmetric theories thus may play important role in nonsupersymmetric theories as well. MAC criterion, though not used as the guiding principle, 
supports our ideas, as it can be shown easily that among the bifermion channels, the ones considered here  ($ \phi^{i A} $ and  $  {\tilde \phi}^{i}_j $)
are the two most attractive channels.

 \section{NonAbelian monopoles and QCD:  ten lessons }

To conclude, let us draw some lessons from these discussions. 

\begin{description}
  \item[(i)]  Confinement mechanism (probably very) different from  the na\"{i}ve  dual superconductor mechanism might be at work in the real world of strong interactions.
  \item[(ii)]  A YM gauge system can generate a dynamically richer IR  effective system, with e.g.,  a larger number of  degrees of freedom, typically involving nonAbelian monopoles and dyons, than expected from an Abelian scenario (e.g.,  $SU(N) \to U(1)^{N-1}$).
  \item[(iii)]   Flavor symmetry, in vacua with a color-flavor locked symmetry,  prevents the YM gauge interactions from Abelianizing dynamically \footnote{This occurred also in the $\psi \chi \eta$ model of Section~\ref{psichieta}.}.
\item[(iv)]      This typically occurs in models where nonAbelian vortices are produced at low energies. Through the monopole-vortex correspondence by topology and stability, such a system can produce nonAbelian monopoles as endpoints of the nonAbelian vortices. These monopoles carry nonAbelian, continuous moduli, free from the classic "difficulties".    

\item[(v)]  The endpoint monopoles -  the source and sink of the nonAbelian vortex - carry the charges of the dual group, ${\tilde H}$, even if they arise from the gauge symmetry breaking
\be   G \to H \to \emptyset\;
\ee
where ${\tilde H}$ is the dual of  $H$ (see Table~\ref{GNOduals}), in accordance with the GNO quantization rule \cite{GNO}.

  \item[(vi)]   Recent series of work on the $CP^{N-1}$ model on finite-width ($L$)  $2D$ worldsheet, establishes that  no dynamical breaking of the 
   isometry group $SU(N)$ occurs even for a finite string.  The system has a unique phase, sometimes called confinement phase, which smoothly goes over   in the $L\to \infty$ limit to the known physics of the $2D$ $CP^{N-1}$ model.
  \item[(vii)]    This amounts to a statement that the nonAbelian monopoles as found in  (iv)  are indeed quantum mechanical objects, carrying fluctuating $CP^{N-1}$ moduli,  and transforming  as in ${\underline N}$ of the new, dual $SU(N)$ group.  
 
  \item[(viii)]   The fact that the $CP^{N-1}$  fluctuations become strongly coupled in the large-distance limit,  and the ground state is a unique string, instead of 
  classically continuously degenerate $CP^{N-1}$ vacua, is perfectly consistent with the (electromagnetic-type) duality. The $H$ theory being in a Higgs phase,  the dual 
  ${\tilde H}$ theory is in confinement phase.

  \item[(ix)]   In ordinary QCD
 a nonAbelian scenario     
   \be    SU(3) \to \frac{SU(2) \times U(1)}{{\mathbf Z}_2}\;, \label{nonAbreakingBis}
  \ee
 could imply a complicated, strongly-coupled nonlocal infrared-fixed-point, lying hidden nearby the confining vacuum we live in.  From the experiences in ${\cal N}=2$ theories, 
 and from the way one understands the quantum nonAbelian monopoles, it is likely that the nonAbelian QCD vacuum  (\ref{nonAbreakingBis})
 crucially depends on the existence in Nature of the two light flavors:  $u$ and $d$ quarks. Confinement and chiral symmetry realization could be related subtly by the condensation of nonAbelian monopoles or dyons carrying flavor quantum numbers.
 
 \item[(x)]  Interesting examples of RG flow with confining vacua "nearby"  such strongly-coupled IRFP conformal theories are found in the context of softly broken
   ${\cal N}=2$  SQCD.   
    There seems to be some analogy between these systems and the real-world QCD. 
  
\end{description}

\section*{Ackowledgments}

K.K. thanks {$CP^3$} Origins, USD, for inviting him to deliver a  {$CP^3$} Origins distinguished lecture, providing him with an excellent opportunity to review some older and latest ideas on monopoles, vortices and confinement. 
 His gratitude goes to all of his collaborators.  
 K.K.  thanks  Stefano Bolognesi, Sven Bjarke Gudnason and Misha Shifman for useful comments.

\end{document}